\RequirePackage[switch,columnwise]{lineno}
\documentclass[showpacs,twocolumn,preprintnumbers,showkeys,superscriptaddress,amsmath,amssymb,nofootinbib]{revtex4-2}
\usepackage{bbold}
\usepackage{color}
\usepackage{latexsym}
\usepackage{amsmath}
\usepackage{amssymb}
\usepackage{epstopdf}
\usepackage{epsfig}
\usepackage[utf8]{inputenc}
\usepackage{amsfonts}
\usepackage{bm}
\usepackage{bbold}
\usepackage{amssymb}

\usepackage{eufrak}
\usepackage{euscript}
\usepackage{pstricks}
\usepackage{graphics}
\usepackage{graphicx}
\usepackage{picture}

\newcommand{\be}{\begin{equation}}
\newcommand{\ee}{\end{equation}}
\newcommand{\ba}{\begin{eqnarray}}
\newcommand{\ea}{\end{eqnarray}}

\def\ni{\noindent}

\begin{document}

%\begin{center}

%
\title{\Large Considerations on the modified Maxwell electrodynamics in the presence of an electric and magnetic background}
%

%\bigskip
%\bigskip

\author{M. J. Neves} \email{mariojr@ufrrj.br}
%\affiliation{Department of Physics and Astronomy, University of Alabama, Tuscaloosa, Alabama 35487, USA}
\affiliation{Departamento de F\'{i}sica, Universidade Federal Rural do Rio de Janeiro, BR 465-07, 23890-971, Serop\'edica, RJ, Brazil}

\author{Patricio Gaete} \email{patricio.gaete@usm.cl}
\affiliation{Departamento de F\'{i}sica and Centro Cient\'{i}fico-Tecnol\'ogico de Valpara\'{i}so-CCTVal,
Universidad T\'{e}cnica Federico Santa Mar\'{i}a, Valpara\'{i}so, Chile}

\author{ L. P. R. Ospedal }  \email{leonardo.ospedal@ufrgs.br}
 \affiliation{ Instituto de F\'isica, Universidade Federal do Rio Grande do Sul,
Av. Bento Gon\c{c}alves 9500, CEP 91501-970, Porto Alegre, Brazil}

\author{J. A. Helay\"el-Neto}\email{helayel@cbpf.br}
\affiliation{Centro Brasileiro de Pesquisas F\'isicas, Rua Dr. Xavier Sigaud
150, Urca, Rio de Janeiro, Brazil, CEP 22290-180}

%%%%%%%%%%%%%%%%%%%%%%%%%%%%%%%%%%%%%%%%%%%%%%%%%%%%%%%%%%%%%%%%%%%%%%%%%%%%%%%%%%%%%%%%%%%%%%%%%%%%%%%%%%%%%%%%%%%%%%%%%%%

\date{\today}

\begin{abstract}
\ni

The properties of the modified Maxwell electrodynamics (ModMax) are investigated in presence
of external and uniform electric and magnetic fields. We expand the non-linear theory around an
electromagnetic background up to second order in the propagating fields to obtain the permittivity and
permeability tensors, dispersion relations, group velocity and refractive indices as
functions of external fields. The case with perpendicular background fields is contemplated. The phenomenon of birefringence is discussed and the fundamental role of the ModMax parameter becomes clear. We calculate the difference of the refractive indices in terms of
this parameter and the external fields. Finally, we set up a scenario where the axion is present and compute the interaction energy for the coupled axion-ModMax electrodynamics if a magnetic background field is considered. This calculation is carried out in the framework of the gauge-invariant, but path-dependent variables formalism. Our results show that the interaction energy contains a linear component, leading to the confinement of static probe charges where the interference between the ModMax parameter, the axion mass and the axion-photon coupling constant is pointed out.

\end{abstract}

%\pacs{11.15.-q; 11.10.Ef; 11.10.Nx}

%\keywords{ModMax electrodynamics, conformal field theory, non-linear electrodynamics.}

\maketitle

\pagestyle{myheadings}
\markright{Considerations on the modified Maxwell electrodynamics in the presence of an electric and magnetic background}

%%%%%%%%%%%%%%%%%%%%%%%%%%%%%%%%%%%%%%%%%%%%%%%%%%%%%%%%%%%%%%%%%%%%%%%%%%%%%%%%%%%%%%%%%%%%%%%%%%%%%%%%%%%%%%%%%%%%%%%%%%%%%%%%%%%%%%%%%%%%%%%%%%%%%%%%%%%%%%%%%%%%%%%%%%%%%%%%%%%%%%%%%%%%%%
%\newpage

%
\section{Introduction}
As widely-known, non-linear electrodynamics has a long history, beginning from the pioneering paper by Born and Infeld (BI) \cite{BornI}, who introduced their theory in order to overcome the intrinsic divergences in the Maxwell theory, at short distances. In passing, we mention that, just like Maxwell electrodynamics, the BI electrodynamics displays no vacuum birefringence. We also recall here that, after the development of Quantum Electrodynamics (QED) \cite{Fermi1,Fock,Fermi2,Dirac}, Heisenberg and Euler (HE) \cite{HEuler} proposed a new non-linear effective theory by summing up the quantum effects of virtual electrons and positrons. Indeed, this new theory contains a striking prediction of the QED, that is, the light-by-light scattering arising from the interaction of photons with virtual electron-positron pairs. It should be further noted that one of the most interesting physical consequences of the HE result is vacuum birefringence. In other words, when the quantum vacuum is stressed by external electromagnetic fields, it behaves like a birefringent material medium. Let us also mention that this physical effect has been emphasized from different viewpoints \cite{Adler, Pistoni, Ruffini, Dunne, Rizzo, Sarazin}. In the context of non-linear electrodynamics, all the solutions to the  non-birefringence condition are shown in ref. \cite{Russo2022}. Nevertheless, this optical phenomenon has not yet been confirmed \cite{Bamber,Burke,Tommasini,Ejlli}.
It is noteworthy to recall here that a few years ago the ATLAS and CMS Collaborations at the Large Hadron Collider (LHC) have reported on the high-energy photon-photon pair emission from virtual photon-photon scattering in ultraperipheral Pb-Pb collisions \cite{ATLAS,CMS}. It is remarkable to notice that, in these results, there is no modification of the optical properties of the vacuum \cite{Robertson}. Besides, the new era of high-power LASER facilities also provides us conditions to probe quantum vacuum non-linearities \cite{Battesti, Ataman}. Among various experiments, a promising proposal is the DeLLight Project \cite{Couchot}, based on the induced change in the vacuum refractive index by virtue of non-linearities in electrodynamics.
%{\color{red} From the theoretical viewpoint, we cite the motivations :
%(i) exact Black-hole solutions motivates the non-linear electrodynamics coupled to gravity \cite{Maceda21,Bordo21,Amirabi21,Kruglov2022,Barrientos22,Ali22}.
%(ii) The TT-like deformation provides the emergence of the non-linear electrodynamics \cite{Velni22,Yekta,Conti22,Ferko22}.
%}
%

%
In the spirit of the foregoing remarks, we have considered the physical effects presented by different models of $(3+1)$-D non-linear electrodynamics in vacuum \cite{MJNevesPRD2021,GEN_BI,LOG}. This has led us to fresh insights on quantum vacuum non-linearities in different contexts. For instance, in the generalized Born-Infeld and Logarithmic electrodynamics, our analysis reveals that the field energy of a point-like charge is finite, apart from displaying the vacuum birefringence phenomenon. Additionally, we have examined the lowest-order modifications of the static potential within the framework of the gauge-invariant but path-dependent variables formalism, which is an alternative to the Wilson loop approach \cite{Gaete97}.
Moreover, a novel non-linear modification of Maxwell electrodynamics that preserves all its symmetries including the electric-magnetic duality and the conformal invariance has recently been proposed \cite{Sorokin1,Sorokin2,Sorokin3}, which is referred to as ModMax electrodynamics. Certainly, the interest in studying non-linear electrodynamics is mainly due to its potentially significant contributions to light-by-light scattering and in the description of novel black hole solutions (see, for instance, refs. \cite{Maceda21,Bordo21,Amirabi21,Kruglov2022,Barrientos22,Ali22}). It is of relevance to point out that the ModMax model can be obtained from $ T \, \overline{T}$-like deformations of Maxwell electrodynamics \cite{Velni22,Yekta,Conti22,Ferko22}, and also within the context of string theory \cite{Nastase2022}.

%Certainly, the interest in studying nonlinear electrodynamics is mainly due to its potentially significant contributions to light-by-light scattering and in the description of certain materials in condensed matter physics \cite{CondMatter}.
%

%
From the previous considerations, and given the interest and importance related to photon-photon interaction physics, this work sets out to further elaborate on a number of phenomenological consequences presented by the ModMax electrodynamics. More specifically, we shall focus our attention on the birefringence, dispersion relations, as well as the computation of the static potential along the lines of the refs. \cite{GEN_BI,LOG,Gaete_AHEP_2021,Gaete_EPJC_2022}. We shall present results of the ModMax ED as a classical field theory and the continuity equation for the energy-momentum tensor is derived . %Using the result of the energy-momentum tensor, we also obtain the spatial components, that are known as the ModMax stress tensor, and posteriorly, the general angular momentum
We also work out the ModMax stress tensor and its angular momentum tensor split into its orbital and spin components. The electric/magnetic properties of the vacuum as a material medium ruled by the ModMax ED are studied in presence of uniform electric and magnetic backgrounds.
%

%
%The particular case in which the electric background is perpendicular to the magnetic one
We focus on the particular case with perpendicular electric and magnetic backgrounds
because the CP-symmetry of the theory is preserved under this condition. Bearing this in mind, we study the wave propagation effects in the linearized ModMax ED. We obtain the permittivity and permeability tensors, the dispersion relations, the refractive index and  group velocity as functions of the ModMax parameter, and the external electric and magnetic fields.
The phenomenon of birefringence in the electromagnetic background is studied in presence of an external magnetic field. Some particular cases are discussed and the recent results in the literature of ModMax ED are correctly recovered. As expected, all the results of the Maxwell ED are also correctly reproduced whenever the ModMax parameter goes to zero.
In the final part of the paper, we present our motivations to introduce ModMax ED coupled to an axion scalar field through an axion-(ModMax)photon interaction term. In this case, we do not consider an electric background, and focus on the axionic ModMax ED in presence of a magnetic background field. In so considering, we study the confinement properties of this new model by computing the interaction energy for a pair of static probe charges within the framework of gauge-invariant but path-dependent variables formalism. In particular, we shall be interested in the dependence of the confinement properties interms of the axion mass, axion-photon coupling and the ModMax parameter.
Our contribution is organized as follows : In Section \ref{sec2}, we quickly review the setup of ModMax theory, derive the field equations and conservation laws. In Section \ref{sec3}, we introduce the prescription of a uniform electromagnetic background in order to expand the ModMax ED up to second order in the propagating fields. In Section \ref{sec4}, the properties of the wave propagation (dispersion relations, refractive index, and group velocity) are calculated in the situation where the electric and magnetic background fields are perpendicular. In Section \ref{sec5}, we investigate the phenomenon of birefringence in the presence of electromagnetic background fields. Subsequently, in Section \ref{sec6}, we consider axionic ModMax electrodynamics and we show that the interaction energy contains a linear potential, leading to the confinement of static probe charges. Finally, our conclusions are cast in Section \ref{sec7}.

At this point, we place a caveat on the fact that Sections III, IV and V deal essentially with the inspection of material properties of the vacuum in presence of external electric and magnetic fields, by considering the linearized approximation in the propagating fields. To our sense, it is important to make clear to the reader that this issue has been fairly-well investigated and discussed in a series of pioneer papers, as quoted in the works cast in refs. \cite{Plebanski68,Birula70,Boillat70,Boillat72,Bialynicki83,Denisov17}. Nevertheless, in the above mentioned Sections III, IV and V, we endeavor to propose an extended re-assessment of the aforementioned classical papers while choosing the particular ModMax ED as our object of investigation to illustrate the propagation of the linear waves. We emphasize however that our attempt is not a mere review. We work out new results, such as the explicit expressions for refractive indices and group velocities, conditions to establish dichroism of the vacuum, find conditions that involve both the ModMax $\gamma$-parameter and the external fields to inspect the positivity of the permittivity and permeability tensors in the special case where the external (electric and magnetic) fields are orthogonal to each other. We would like to recall that in the paper \cite{Sorokin1}, the authors consider the specific situation of (anti) parallel external fields. Moreover, we also present a discussion of the angular momentum tensor and the corresponding spin vector for ModMax ED.

We adopt the natural units $\hbar=c=1$  with $4 \pi \epsilon_0 = 1$, and the Minkowski
metric $\eta^{\mu\nu}=\mbox{diag}(+1,-1,-1,-1)$. The electric and magnetic fields have squared-energy mass dimension in which the conversion
of Volt/m and Tesla (T) to the natural system is as follows: $1 \, \mbox{Volt/m}=2.27 \times 10^{-24} \, \mbox{GeV}^2$ and $1 \, \mbox{T} =  6.8 \times 10^{-16} \, \mbox{GeV}^2$, respectively.

%%%%%%%%%%%%%%%%%%%%%%%%%%%%%%%%%%%%%%%%%%%%%%%%%%%%%%%%%%%%%%%%%%%%%%%%%%%%%%%%%%%%%%%%%%%%%%%%%%%%%%%%%%%%%%%%%%%%%%%%%%%%%%%%%%%%%%%%%%%%%%%%%%%%%%%%%%%%%%%%%%%%%%%%%%%%%%%%%%%%%%%%%%%%%%%%%%%%%%%%%%%%%%%%%%%%%%%%%%%%%%%%%%%%%%%%%%%%%%%%%%%%%%%%%%%%%%%%%%%%%%%%%%%%%%%%%%%%%%%%%%%%%%%%%%%%%%%%%%%%%%%%%%%%%%%%%%%%%%%%%%%%%%%%%%%%%%%%%%%%%%%%%%%%%%%%%%%%%%%%%%%%%%%%%%%%%%%%%%%%%%%%%%%%%%%%%%%%%%%%%%%%%%%%%%%%%%%%%%%%%%%%%%%%%%%%%%%%%%%%%%%%%%%%%%%%%%%%%%%%%%%%%%%%%%%%%%%%%%%%%%%%%%%%%%%%%%%%%%%%%%%%%%%%%%%%%%%%%%%%%%%%%%%%%%%%%%%%%%%%%%%%%%%%%%%%%%%%%%%%%%%%%%%%%%%%%%%%%%%%%%%%%%%%%%%%%%%%%%%%%%%%%%%%%%%%%%%%%%%%%%%%%%%%%%%%%%%%%%%%%%%%%%%%%%%%%%%%%%%%%%%%%%%%%%%%%%%%%%%%%%%%%%%%%%%%%%%%%%%%%%%%%%%%%%%%%%%%%%%%%%%%%%%%%%%%%%%%%%%%%%%%%%%%
%

%
\section{The modified Maxwell electrodynamics}
\label{sec2}
We start off the classical description of the ModMax ED through the Lagrangian density:
\begin{eqnarray}\label{ModMaxL}
{\cal L}_{MM}=\cosh\gamma \, {\cal F} + \sinh\gamma \sqrt{{\cal F}^2+{\cal G}^2}+J_{\mu}A^{\mu} \; ,
\end{eqnarray}
where ${\cal F}$ and ${\cal G}$ denote the Lorentz and gauge invariants,
\begin{subequations}
\begin{eqnarray}
{\cal F}\!&=&\!-\frac{1}{4} \, F_{\mu\nu}^{2}=\frac{1}{2} \, \left( \, {\bf E}^2-{\bf B}^2 \, \right) \; ,
\label{invF}
\\
%\hspace{0.2cm} \mbox{and} \hspace{0.2cm}
{\cal G}\!&=&\!-\frac{1}{4} \, F_{\mu\nu}\widetilde{F}^{\mu\nu}={\bf E}\cdot{\bf B} \; ,
\label{invG}
\end{eqnarray}
\end{subequations}
with $\gamma$ being a real parameter, that satisfies the condition $\gamma \geq 0$ to insure the causality
and unitarity of the model \cite{Sorokin1}, and $J^{\mu}=(\rho,{\bf J})$ is a classical source of charge and current densities. The $F^{\mu\nu}=\partial^{\mu}A^{\nu}-\partial^{\nu}A^{\mu}=\left( \, -E^{i} \, , \, -\epsilon^{ijk} B^{k} \, \right)$
is the skew-symmetric field strength tensor, and $\widetilde{F}^{\mu\nu}=\epsilon^{\mu\nu\alpha\beta}F_{\alpha\beta}/2=\left( \, -B^{i} \, , \, \epsilon^{ijk} E^{k} \, \right)$ corresponds to the dual tensor. The usual Maxwell electrodynamics is recovered when $\gamma \rightarrow 0$.
The action principle yields the field equations
\begin{equation}\label{eq}
\partial_{\mu}\left[ \, \cosh\gamma \, F^{\mu\nu}+\sinh\gamma \, \frac{F^{\mu\nu}\,{\cal F}+\widetilde{F}^{\mu\nu}\,{\cal G}}{\sqrt{{\cal F}^2+{\cal G}^2}} \, \right]=J^{\nu} \; ,
\end{equation}
and the dual tensor satisfies the Bianchi identity $\partial_{\mu}\widetilde{F}^{\mu\nu}=0$. The charge conservation obeys the continuity equation $\partial_{\mu}J^{\mu}=0$, as in the Maxwell ED. The equations in terms of the electric and magnetic fields can be written as :
\begin{subequations}
%\begin{widetext}
\begin{eqnarray}
\nabla\cdot{\bf D} \!&=&\! \rho
\hspace{0.3cm} , \hspace{0.3cm}
%-\frac{d_{3}}{c_{1}} \, c\, {\bf B} \cdot \nabla\left( {\bf B}\cdot{\bf b} \right) =0 \; ,
%\hspace{0.3cm}
\nabla\times{\bf E}+\partial_{t}{\bf B}={\bf 0} \; ,
\label{EqDivD}
\\
\nabla\cdot{\bf B} \!&=&\! 0
\hspace{0.3cm} , \hspace{0.3cm}
\nabla\times{\bf H}-\partial_{t}{\bf D}={\bf J} \; ,
\label{EqrotH}
\hspace{0.5cm}
\end{eqnarray}
%\end{widetext}
\end{subequations}
where ${\bf D}$ and ${\bf H}$ are defined by
\begin{subequations}
\begin{eqnarray}
{\bf D} = \cosh\gamma \, {\bf E} + \sinh\gamma \, \frac{{\cal F} \, {\bf E}+{\cal G} \, {\bf B}}{\sqrt{{\cal F}^2+{\cal G}^2}} \; ,
\label{D}
\\
{\bf H} = \cosh\gamma \, {\bf B} + \sinh\gamma \, \frac{{\cal F} \, {\bf B}-{\cal G} \, {\bf E}}{\sqrt{{\cal F}^2+{\cal G}^2}} \; .
\label{H}
\end{eqnarray}
\end{subequations}
%

%
%From (\ref{eq}), the external current satisfies the continuity equation $\partial_{\mu}J^{\mu}=0$.
Multiplying the Bianchi identity by $F^{\mu\nu}$, and using the eq. (\ref{eq}),
we arrive at the conservation law
\begin{eqnarray}\label{EqThetaJ}
\partial_{\mu}\Theta^{\mu\nu}=J_{\rho}\,F^{\rho\nu} \; ,
\end{eqnarray}
where the energy-momentum tensor of the ModMax ED is given by
\begin{equation}\label{tensorEnMomentum}
\Theta^{\mu\nu}=
\left( \, F^{\mu\rho} F_{\rho}^{\;\;\nu}
-\eta^{\mu\nu} \,{\cal F}\right)\left[ \cosh\gamma+ \frac{ \sinh\gamma \, {\cal F} }{\sqrt{{\cal F}^2+{\cal G}^2}} \right] \; .
\end{equation}
This tensor has the properties of symmetry in the $(\mu \leftrightarrow \nu)$ indices and is gauge invariance. In the case of the ModMax in a free space
(with no charge and current densities), the expression (\ref{EqThetaJ}) satisfies immediately the continuity equation, $\partial_{\mu}\Theta^{\mu\nu}=0$,
where the $\Theta^{00}$- and $\Theta^{0i}$-components denote the conserved energy and momentum densities stored in the EM fields, namely,
\begin{subequations}
\begin{eqnarray}
\Theta^{00} \!\!&=&\!\! \frac{1}{2} \, ({\bf E}^{\,2}\!+\!{\bf B}^2) \! \left[ \cosh\gamma \!+\! \frac{ \sinh\gamma \, ({\bf E}^2-{\bf B}^2) }{\sqrt{ ({\bf E}^2-{\bf B}^2)^2+4({\bf E}\cdot{\bf B})^{2}}} \right] ,
\label{Thetacomp00}
\nonumber \\
\\
\Theta^{0i} \!\!&=&\!\! \left( {\bf E}\! \times\! {\bf B} \right)^{i} \! \left[ \cosh\gamma \!+\! \frac{ \sinh\gamma \, ({\bf E}^2-{\bf B}^2) }{\sqrt{ ({\bf E}^2-{\bf B}^2)^2+4({\bf E}\cdot{\bf B})^{2}}} \right] .
\label{Thetacomp0i}
\end{eqnarray}
\end{subequations}
Notice that both components in eqs. (\ref{Thetacomp00}) and (\ref{Thetacomp0i}) are not defined if the electric and magnetic fields, for example, satisfy the relation of null electromagnetic fields, ${\bf E}^2-{\bf B}^2=0$ and ${\bf E}\cdot{\bf B}=0$. To remove this singularity, it is convenient to define the Legendre transformation :
\begin{eqnarray}
{\cal H}({\bf D},{\bf B})={\bf E} \cdot {\bf D} - {\cal L}_{MM}({\bf E},{\bf B}) \; ,
\end{eqnarray}
and using the ModMax Lagrangian (\ref{ModMaxL}), the Hamiltonian density turns out to be given by
\begin{eqnarray}\label{Hmm}
{\cal H}_{MM} \!\!&=&\!\! \frac{1}{2} \, \cosh\gamma \left({\bf D}^2+{\bf B}^2\right)
\nonumber \\
&&
\hspace{-0.5cm}
-\frac{1}{2} \, \sinh\gamma \, \sqrt{ \left({\bf D}^2-{\bf B}^2\right)^{2}+4\left({\bf D}\cdot{\bf B}\right)^{2}} \; ,
\end{eqnarray}
which is positive-definite if the ModMax parameter obeys the condition
\begin{eqnarray}
\tanh\gamma \, < \, \sqrt{ \, \frac{({\bf D}^2+{\bf B}^2)^{2}}{({\bf D}^2-{\bf B}^2)^{2}+4({\bf D}\cdot{\bf B})^{2}} \, } \; .
\end{eqnarray}
The quantity (\ref{Hmm}) is well-defined for any ${\bf D}$ and ${\bf B}$. Furthermore, it is manifestly invariant under the duality transformations ${\bf D}^{\prime}+i\,{\bf B}^{\prime}=e^{i\alpha} \, ({\bf D}+i\,{\bf B})$, with $\alpha$ being a real parameter. The Poynting vector, given by the $\Theta^{0i}$-component in terms of the ${\bf D}$- and ${\bf B}$- read as below:
\begin{eqnarray}
{\bf S}_{P}={\bf D} \times {\bf B} \; .
\end{eqnarray}
The spatial component $\nu=j$ in the conservation law (\ref{EqThetaJ}) yields the expression
\begin{eqnarray}
\nabla\cdot \overleftrightarrow{{\bf T}}-\partial_{t}{\bf S}_{P}={\bf f}_{L} \; ,
\end{eqnarray}
where ${\bf f}_{L}=\rho \, {\bf E} \,+\, {\bf J}\times{\bf B}$ is the Lorentz force density,
while the components of the Maxwell stress tensor $\overleftrightarrow{{\bf T}}$
are given by
\begin{eqnarray}
T^{ij} \!&=&\! \left[ \, \cosh\gamma+\frac{ \sinh\gamma \, {\cal F} }{ \sqrt{ {\cal F}^2+{\cal G}^2 } } \, \right] \times
\nonumber \\
&&
\times \left[ \, E^{i}\,E^{j}+B^{i}\,B^{j}-\delta^{ij} \, \frac{1}{2} \, ({\bf E}^2+{\bf B}^2) \, \right]
\; .
\end{eqnarray}
The Lorentz force has the same definition as in Maxwell ED, namely, it is obtained
by integrating the corresponding density over a region of the space.
%

%
%Above, in eq. (\ref{tensorEnMomentum}), we give the expression for the symmetric and gauge-invariant
Having established the energy-momentum tensor of the
ModMax ED, we now proceed to the angular momentum tensor. Let us adopt the canonical approach and split it into an orbital (OAM) and spin (SAM) components as follows:
\begin{eqnarray}
M^{\mu}_{\;\;\,\alpha\beta}=x_{\alpha}\,T^{\mu}_{\;\;\,\beta}-x_{\beta}\,T^{\mu}_{\;\;\,\alpha}+S^{\mu}_{\;\;\,\alpha\beta}  \; ,
\end{eqnarray}
where $T^{\mu}_{\;\;\,\beta}$ stands for the canonical energy-momentum tensor, whereas $S^{\mu}_{\;\;\,\alpha\beta}$ expresses the spin contribution:
\begin{equation}
T^{\mu}_{\;\;\,\beta}=\left[ \cosh\gamma\, F^{\mu\lambda}\!+\!\sinh\gamma \, \frac{ {\cal F} \, F^{\mu\lambda}+{\cal G} \, \tilde{F}^{\mu\lambda} }{\sqrt{ {\cal F}^2+{\cal G}^2 }} \right]\!\partial_{\beta}A_{\lambda}-\delta^{\mu}_{\;\,\,\beta}\, {\cal L}_{MM} \, , \;\;
\end{equation}
and
\begin{eqnarray}
S^{\mu}_{\;\;\,\alpha\beta} \!\!&=&\!\! \left[ \cosh\gamma+ \frac{ \sinh\gamma \, {\cal F}}{\sqrt{{\cal F}^2+{\cal G}^2}} \right](F^{\mu}_{\;\;\,\beta}A_{\alpha}-F^{\mu}_{\;\;\,\alpha}A_{\beta})
\nonumber \\
&&
\hspace{-0.5cm}
+ \frac{\sinh\gamma \, {\cal G}}{\sqrt{{\cal F}^2+{\cal G}^2}} \, (\widetilde{F}^{\mu}_{\;\;\,\beta}A_{\alpha}-\widetilde{F}^{\mu}_{\;\;\,\alpha}A_{\beta})
 \; .
\end{eqnarray}
Consequently, the corresponding spin vector can be read from the expression below:
\begin{eqnarray}
{\bf S}_{spin} \!\!&=&\!\! \int d^{3}{\bf x} \, \left\{
\left[ \cosh\gamma+ \frac{ \sinh\gamma \, {\cal F}}{\sqrt{{\cal F}^2+{\cal G}^2}} \right] ({\bf E}\times{\bf A})
\right.
\nonumber \\
&&
\hspace{-0.5cm}
\left.
+ \frac{ \sinh\gamma \, {\cal G}}{\sqrt{{\cal F}^2+{\cal G}^2}} \, ({\bf B}\times{\bf A})
\right\} \; ,
\end{eqnarray}
which can be recast as
%In terms of ${\bf D}$, the spin vector is
%
\begin{eqnarray}
{\bf S}_{spin} = \int d^{3}{\bf x} \; {\bf D} \times {\bf A} \; .
\end{eqnarray}
Both the OAM and SAM components are not gauge-invariant and
a physically unambiguous splitting into these two pieces is still
controversial and object of debate. Actually, the formal separation
into orbital and spin parts of an optical field first appeared in a paper
by J. Humblet \cite{Humblet}. In 1932, C. G. Darwin pioneered the investigation of the angular momentum tensor of
electromagnetic radiation, though he did not exploit the OAM-SAM splitting \cite{Darwin}. For an updated discussion, we address the interested readers to the papers \cite{Marrucci,Stewart}, where alternative decompositions into OAM and SAM components are presented and discussed with a great deal of details.

\section{The linearized ModMax electrodynamics in an EM background}
\label{sec3}
The ModMax ED can be linearized by expanding the $A^{\mu}$-potential around
a uniform and constant EM field, as $A_{\mu}=a_{\mu}+A_{0\mu}$, where $a_{\mu}$
is the excitation interpreted as the photon field, whereas $A_{0\mu}$ is the potential associated
with the EM background. We adopt a similar procedure and notations of ref. \cite{MJNevesPRD2021}.
The expansion is considered up to second order in the photon fluctuation $a^{\mu}$. Thereby, the tensor $F_{\mu\nu}$ is also decomposed as $F_{\mu\nu}=f_{\mu\nu}+F_{0\mu\nu}$, in which $f^{\mu\nu}=\partial^{\mu}a^{\nu}-\partial^{\nu}a^{\mu}=\left( \, -e^{i} \, , \, -\epsilon^{ijk} \, b^{k} \, \right)$ is the EM field strength tensor of the propagating field, and $F_{0}^{\;\,\mu\nu}=\partial^{\mu}A_{0}^{\;\,\nu}-\partial^{\nu}A_{0}^{\;\,\mu} =\left( \, -E_{0}^{\,\,i} \, , \, -\epsilon^{ijk} \, B_{0}^{\,\,k} \, \right)$ is the field strength associated with the background electric and magnetic fields. Now, by expanding the Lagrangian density (\ref{ModMaxL}) around the external field, we arrive at
\begin{eqnarray}\label{LMM2}
{\cal L}_{MM}^{(2)} \!\!&=&\!\! -\frac{1}{4} \, c_{1} \, f_{\mu\nu}^{\, 2}
-\frac{1}{4} \, c_{2} \, f_{\mu\nu} \, \widetilde{f}^{\mu\nu}
-\frac{1}{2} \, G_{0\mu\nu} \, f^{\mu\nu}
\nonumber \\
&&
\hspace{-0.4cm}
+\,\frac{1}{8} \, Q_{0\mu\nu\kappa\lambda} \, f^{\mu\nu} \, f^{\kappa\lambda} \; ,
\end{eqnarray}
where $G_{0\mu\nu}=c_{1} \, F_{0\mu\nu}+ c_{2} \, \widetilde{F}_{0\mu\nu}$ and
$Q_{0\mu\nu\kappa\lambda}=d_{1} \, F_{0\mu\nu}F_{0\kappa\lambda}
+d_{2} \, \widetilde{F}_{0\mu\nu}\widetilde{F}_{0\kappa\lambda}
+d_{3} \, F_{0\mu\nu}\widetilde{F}_{0\kappa\lambda}
+ d_{3} \, \widetilde{F}_{0\mu\nu} F_{0\kappa\lambda}$ are tensors
that depend on the components of the EM background fields.
The $\widetilde{f}^{\mu\nu}=\epsilon^{\mu\nu\alpha\beta}f_{\alpha\beta}/2=\left( \, -b^{i} \, , \, \epsilon^{ijk} e^{k} \, \right)$
and $\widetilde{F}_{0}^{\;\,\mu\nu}=\epsilon^{\mu\nu\alpha\beta}F_{0\alpha\beta}/2=\left( \, -B_{0}^{\;\,i} \, , \, \epsilon^{ijk}
E_{0}^{\;\,k} \, \right)$ are the dual field strength tensors of the propagating and background fields, respectively.
The $\widetilde{f}^{\mu\nu}$ satisfies the Bianchi identity $\partial_{\mu}\widetilde{f}^{\mu\nu}=0$ from which
there emerge the homogeneous field equations:
\begin{equation} \label{hom_eq}
\nabla \cdot {\bf b} = 0 \, \; , \, \; \nabla \times {\bf e} = - \partial_t {\bf b} \, .
\end{equation}
The coefficients $c_{1}$, $c_{2}$, $d_{1}$, $d_{2}$ and $d_{3}$ of this expansion are defined as cast below:
\begin{eqnarray}\label{coefficients}
c_{1}=\left.\frac{\partial{\cal L}_{MM}}{\partial{\cal F}}\right|_{{\bf E}_{0},{\bf B}_{0}}
\, , \,
\left. c_{2}=\frac{\partial{\cal L}_{MM}}{\partial{\cal G}}\right|_{{\bf E}_{0},{\bf B}_{0}}
\, , \,
\nonumber \\
\left. d_{1}=\frac{\partial^2{\cal L}_{MM}}{\partial{\cal F}^2}\right|_{{\bf E}_{0},{\bf B}_{0}}
\, , \,
\left. d_{2}=\frac{\partial^2{\cal L}_{MM}}{\partial{\cal G}^2}\right|_{{\bf E}_{0},{\bf B}_{0}}
\, , \,
\nonumber \\
\left. d_{3}=\frac{\partial^2{\cal L}_{MM}}{\partial{\cal F}\partial{\cal G}}\right|_{{\bf E}_{0},{\bf B}_{0}} ,
\hspace{0.4cm}
\end{eqnarray}
that also depend only on the EM background. Substituting the ModMax Lagrangian density (\ref{ModMaxL}), we obtain :
\begin{eqnarray}\label{coefficientsMM}
c_{1} \!\!&=&\!\!\cosh\gamma+\frac{\sinh\gamma\,{\cal F}_{0}}{\sqrt{{\cal F}_{0}^2+{\cal G}_{0}^2}}
\; , \;
c_{2}=\frac{\sinh\gamma\,{\cal G}_{0}}{\sqrt{{\cal F}_{0}^2+{\cal G}_{0}^2}} \; ,
\nonumber \\
d_{1} \!\!&=&\!\! \frac{\sinh\gamma}{\sqrt{{\cal F}_{0}^2+{\cal G}_{0}^2}} \frac{{\cal G}_{0}^{2}}{{\cal F}_{0}^2+{\cal G}_{0}^2} \; ,
\nonumber \\
d_{2} \!\!&=&\!\! \frac{\sinh\gamma}{\sqrt{{\cal F}_{0}^2+{\cal G}_{0}^2}} \frac{{\cal F}_{0}^{2}}{{\cal F}_{0}^2+{\cal G}_{0}^2} \; ,
\nonumber \\
d_{3} \!\!&=&\!\! \frac{\sinh\gamma}{ \sqrt{ {\cal F}_{0}^2+{\cal G}_{0}^2} } \, \frac{{\cal F}_{0} \, {\cal G}_{0}}{{\cal F}_{0}^2+{\cal G}_{0}^2} \; ,
\end{eqnarray}
in which ${\cal F}_{0}=({\bf E}_{0}^2-{\bf B}_{0}^{2})/2$ and ${\cal G}_{0}={\bf E}_{0}\cdot{\bf B}_{0}$ are the gauge and Lorentz
invariants associated with the electric ${\bf E}_{0}$, and magnetic ${\bf B}_{0}$ fields.
The action principle applied to the Lagrangian density (\ref{LMM2}) yields the field equations :
\begin{eqnarray}\label{Eqfmunu}
\partial^\mu \left[ \, c_1 \, f_{\mu \nu} + c_2 \, \widetilde{f}_{\mu \nu} -
\frac{1}{2} \, Q_{0\mu \nu \kappa \lambda} \, f^{\kappa \lambda} \, \right]
= 0 \; .
\end{eqnarray}
The usual Maxwell ED is recovered when the EM background fields are turn-off
and the ModMax parameter goes to zero. The equations (\ref{Eqfmunu}) can be written into the same
form of eqs. (\ref{EqDivD}) and (\ref{EqrotH}), with the auxiliary fields
%
%in which the components of ${\bf D}$ in (\ref{D}),and ${\bf H}$ in (\ref{H}) are linear combinations of the propagating fields ${\bf e}$ and ${\bf b}$
%
\begin{subequations}
\begin{eqnarray}
D_{i} \!&=&\! \varepsilon_{ij}\,e_{j}+\sigma_{ij}\,b_{j}
%+c_{1} \, E_{i} \, \delta^{(3)}({\bf k}) \, \delta(\omega)
\; ,
\label{Di}
\\
H_{i} \!&=&\! -\sigma_{ji}\,e_{j}+(\mu_{ij})^{-1}\,b_{j}
%+c_{1} \, B_{i} \, \delta^{(3)}({\bf k}) \, \delta(\omega)
\; ,
\label{Hi}
\end{eqnarray}
\end{subequations}
where the permittivity symmetric tensor $\varepsilon_{ij}$, the inverse of the permeability symmetric tensor $(\mu_{ij})^{-1}$, and $\sigma_{ij}$ are given in terms of the electric and magnetic background components, as follows
\begin{subequations}
\begin{eqnarray}
\varepsilon_{ij} \!\!&=&\!\! c_{1} \, \delta_{ij} + d_1 \, E_{0i} \, E_{0j}
+ d_2 \, B_{0i} \, B_{0j}
\nonumber \\
&&
+ \, d_{3} \, E_{0i} \, B_{0j} + d_{3} \, B_{0i} \, E_{0j} \; ,
\label{epsilonij}
\\
\sigma_{ij} \!\!&=&\!\! c_{2} \, \delta_{ij}-d_{1} \, E_{0i} \, B_{0j} + d_{2} \, B_{0i} \, E_{0j}
\nonumber \\
&&
+ \, d_{3} \, E_{0i} \, E_{0j}-d_{3} \, B_{0i} \, B_{0j} \; ,
\label{sigmaij}
\\
(\mu_{ij})^{-1} \!\!&=&\!\! c_{1} \, \delta_{ij} - d_1 \, B_{0i} \, B_{0j} - d_2 \, E_{0i} \, E_{0j}
\nonumber \\
&&
-d_{3} \, E_{0i} \, B_{0j} - d_{3} \, B_{0i} \, E_{0j} \; .
\label{muij}
\end{eqnarray}
\end{subequations}
Notice that the case in which $c_{2} \neq 0$ and $d_3 \neq 0$, CP-symmetry is violated in the linearized theory.
To keep CP invariance, we just consider the situation in which the electric and magnetic background fields are mutually orthogonal.

\section{Wave propagation for perpendicular external fields}
\label{sec4}

If we consider the case with ${\bf E}_{0}$ perpendicular to ${\bf B}_{0}$,
the second Lorentz and gauge invariant quantity ${\cal G}_{0}=0$, so that the coefficients from eq. (\ref{coefficients}) are given by
\begin{eqnarray}\label{coefficientsMM}
c_{1}  & = &  \cosh\gamma+\sinh\gamma \, \mbox{sgn}({\cal F}_{0})
\; , \;
d_{2} = \frac{\sinh\gamma}{|{\cal F}_{0}|} \; ,
\nonumber \\
c_{2} & = &  d_{1} = d_3 = 0 \; ,
\end{eqnarray}
in which $\mbox{sgn}({\cal F}_{0})$ is the signal function of ${\cal F}_{0}$,
where $\mbox{sgn}({\cal F}_{0}) = 1$ if $|{\bf E}_{0}| > |{\bf B}_{0}|$,
and $\mbox{sgn}({\cal F}_{0}) = -1$ if $|{\bf B}_{0}| > |{\bf E}_{0}|$.
Under these conditions, the permittivity tensor, (\ref{epsilonij}), and the
permeability tensor related to eq. (\ref{muij}) turn out to be
\begin{subequations}
\begin{eqnarray}
\varepsilon_{ij} \!&=&\! c_1 \, \delta_{ij}+d_{2} \, B_{0i} \, B_{0j} \; ,
\\
\mu_{ij} \!&=&\! \frac{1}{c_{1}} \left[ \, \delta_{ij}+\frac{ d_{E} \, E_{0i} \, E_{0j} }{1-d_E \, {\bf E}_{0}^{2}} \, \right] \; ,
\end{eqnarray}
\end{subequations}
where the coefficient $d_{E}$ is
\begin{eqnarray}
d_{E} = \frac{d_2}{c_1} = \frac{\tanh(\gamma)}{|{\cal F}_{0}|+\tanh(\gamma)\,{\cal F}_{0}} \; ,
\end{eqnarray}
in which
\begin{subequations}
\begin{eqnarray}
d_{E} \!&=&\! \frac{ 1-e^{-2\gamma} }{E_{0}^2-B_{0}^2} \hspace{0.3cm} \mbox{if} \hspace{0.3cm} |{\bf E}_{0}|>|{\bf B}_{0}| \; ,
\label{dEEB}
\\
d_{E} \!&=&\! \frac{ e^{2\gamma}-1 }{ B_{0}^2-E_{0}^2 } \hspace{0.3cm} \mbox{if} \hspace{0.3cm} |{\bf B}_{0}|>|{\bf E}_{0}| \; .
\label{dEBE}
\end{eqnarray}
\end{subequations}
The eigenvalues of the matrices $\varepsilon_{ij}$ and $\mu_{ij}$ are given by:
\begin{subequations}
\begin{eqnarray}
\lambda_{1\varepsilon} \!&=&\! \lambda_{2\varepsilon} = c_1
\; , \;
\lambda_{3\varepsilon} = c_1 + d_{2} \, {\bf B}_{0}^2 \; ,
\\
\lambda_{1\mu} \!&=&\! \lambda_{2\mu} = \frac{1}{c_1}
\; , \;
\lambda_{3\mu} = \frac{1}{c_1-d_{2}\,{\bf E}_{0}^2}  \; ,
\end{eqnarray}
\end{subequations}
the electric permittivity and magnetic permeability are both positive-definite if the eigenvalues satisfy the conditions
$c_{1} > 0$, $1+d_{E}\,{\bf B}_{0}^2 > 0$, and $1-d_{E} \, {\bf E}_{0}^2 > 0$.
We turn now our attention to the dispersion relations. Considering plane wave solutions, ${\bf e}({\bf x},t)={\bf e}_{0} \, e^{i \, \left({\bf k}\cdot{\bf x}-\omega t\right)}$ and ${\bf b}({\bf x},t)={\bf b}_{0} \, e^{i \, \left({\bf k}\cdot{\bf x}-\omega t\right)}$ in eq. (\ref{Eqfmunu}), the wave equation for the components of the electric amplitude $e_{0i}$ is
\begin{eqnarray}\label{EqMatrix}
M_{ij} \, e_{0j}=0 \; .
\end{eqnarray}
The matrix elements $M_{ij}$ have the form
\begin{equation}\label{Mij}
M_{ij}=\left( \omega^2-{\bf k}^2 \right) \delta_{ij}+k_{i} \, k_{j} + u_{i} \, v_{j} \; ,
\end{equation}
where the vectors are defined by $u_{i}=\omega \, B_{0i}-({\bf k} \times {\bf E}_{0})_{i}$ and $v_{i}=d_{E} \, u_{i}$.
The determinant of the $M$-matrix can be calculated and yields the expression
\begin{equation}
\mbox{det}M \!=\!(\omega^2-{\bf k}^2)\left[ \, \omega^2\left(\omega^2-{\bf k}^2+ {\bf u}\cdot{\bf v}\right)
-\left({\bf u}\cdot{\bf k} \right) \left({\bf v}\cdot{\bf k} \right) \, \right] \; , \; \; \;
\end{equation}
The condition $\mbox{det}M=0$ leads to the usual photon dispersion relation $\omega=|{\bf k}|$
as one of the solutions. The other solutions correspond to the roots of the polynomial equation :
%
%The relation between the frequency, $\omega$, and the wave vector, ${\bf k}$, can be written in a matrix form:
%
%
\begin{eqnarray}\label{eqomega4}
\omega^2 \left( \, P \, \omega^2+Q \, \omega+R \, \right)=0 \; ,
\end{eqnarray}
where $P = 1+d_{E} \, {\bf B}_{0}^2$,
$Q = -2\, d_{E} \, {\bf B}_{0} \cdot ({\bf k}\times{\bf E}_{0})$, and
$R = d_{E} \left({\bf k}\times{\bf E}_{0}\right)^2 -{\bf k}^2-d_{E} \, \left({\bf B}_{0}\cdot{\bf k}\right)^2$.
%
%
%\begin{eqnarray}
%P \!\!&=&\!\! 1+\frac{d_1}{c_1} \, \frac{{\bf E}^2}{c^2}+\frac{d_2}{c_1} \, {\bf B}^2 \; ,
%\nonumber \\
%Q \!\!&=&\!\! -2{\bf k}^2+\frac{d_{1}-d_{2}}{c_{1}} \, \frac{{\bf E}^2}{c^2} \, {\bf k}^2
%-\frac{d_1+d_2}{c_1} \, \left( \frac{{\bf E}}{c}\cdot{\bf k} \right)^2
%\nonumber \\
%&&
%\hspace{-0.5cm}
%+\frac{d_1d_2+d_{3}^{2}}{c_1^{2}} \, \frac{{\bf E}^2}{c^2} \left( \frac{{\bf E}}{c} \times {\bf k}\right)^2
%\!\!+\frac{d_{1}-d_{2}}{c_{1}} \, {\bf B}^2 \, {\bf k}^2
%\nonumber \\
%&&
%\hspace{-0.5cm}
%-\frac{d_1+d_2}{c_1} \, \left( {\bf B}\cdot{\bf k}  \right)^2
%+\frac{d_1d_2+d_{3}^{2}}{c_1^{2}} \, {\bf B}^2 \left({\bf B} \times {\bf k}\right)^2 \; ,
%\nonumber \\
%R \!\!&=&\!\! {\bf k}^4-\frac{d_{2}}{c_{1}} \, {\bf k}^2\left( \frac{{\bf E}}{c}\cdot{\bf k}  \right)^2
%+\frac{d_1}{c_1} \, {\bf k}^2 \left( \frac{{\bf E}}{c}\cdot{\bf k}  \right)^2
%\nonumber \\
%&&
%\hspace{-0.5cm}
%+\frac{d_{3}^{2}-d_1d_2}{c_1^{2}} \, \left(\frac{{\bf E}}{c}\cdot{\bf k}\right)^2 \left( \frac{{\bf E}}{c} \times {\bf k}\right)^2
%\!\!-\frac{d_{1}}{c_{1}} \, {\bf k}^2\left( {\bf B}\times{\bf k}  \right)^2
%\nonumber \\
%&&
%\hspace{-0.5cm}
%+\frac{d_2}{c_1} \, {\bf k}^2 \left( {\bf B}\cdot{\bf k}  \right)^2
%+\frac{d_{3}^{2}-d_1d_2}{c_1^{2}} \, \left({\bf B}\cdot{\bf k}\right)^2 \left({\bf B} \times {\bf k}\right)^2
%\; .
%\end{eqnarray}
%
The zeroes of eq. (\ref{eqomega4}) are $\omega=0$, and the non-trivial solutions :
%
%$\omega_{1}^{(\pm)}=\pm \, \omega_{1}({\bf k})$ and $\omega_{2}^{(\pm)}=\pm \, \omega_{2}({\bf k})$, whose frequencies are shown below:
%
\begin{widetext}
\begin{subequations}
\begin{eqnarray}
\omega^{(-)}({\bf k}) \!&=&\! \frac{d_{E}{\bf B}_{0}\cdot({\bf k}\times{\bf E}_{0})-\sqrt{{\bf k}^2-d_{E}\left({\bf k}\times{\bf E}_{0}\right)^{2}-d_{E}^2\left({\bf E}_{0}^2-{\bf B}_{0}^{2} \right)({\bf k}\cdot{\bf B}_{0})^2 + d_{E}\left({\bf k}\cdot{\bf B}_{0}\right)^{2}+d_{E}\,{\bf B}_{0}^2\,{\bf k}^2 } }{1+d_{E}\,{\bf B}_{0}^2} \; ,
\label{omegam} \\
\omega^{(+)}({\bf k}) \!&=&\! \frac{d_{E}{\bf B}_{0}\cdot({\bf k}\times{\bf E}_{0})+\sqrt{{\bf k}^2-d_{E}\left({\bf k}\times{\bf E}_{0}\right)^{2}-d_{E}^2\left({\bf E}_{0}^2-{\bf B}_{0}^{2} \right)({\bf k}\cdot{\bf B}_{0})^2 + d_{E}\left({\bf k}\cdot{\bf B}_{0}\right)^{2}+d_{E}\,{\bf B}_{0}^2\,{\bf k}^2 } }{1+d_{E}\,{\bf B}_{0}^2}   \; .
%\nonumber \\
%\omega_{2}^{(\pm)} \!&=&\! \pm \, c \, \sqrt{ \frac{-Q+\sqrt{Q^2-4 \, P \, R}}{2P} } \; .
\label{omegap}
\end{eqnarray}
\end{subequations}
\end{widetext}
The limit of usual Maxwell ED $(\gamma \rightarrow 0)$ in eqs. (\ref{omegap}) and (\ref{omegam}) yields the roots $\lim_{\gamma \rightarrow 0} \omega^{(\pm)}({\bf k})=\pm \, |{\bf k}|$, so that the photon dispersion relation is recovered. Thereby, we choose the $\omega^{(+)}$ frequency for the analysis of the refractive index of the medium. Notice that the refractive index $n^{(+)}=|{\bf k}|/\omega^{(+)}({\bf k})$ does not depend on the wavelengths, $(\lambda=2\pi/|{\bf k}|)$, so that dispersion does not occur. It is important to remark that the frequencies in eqs. (\ref{omegam}) and (\ref{omegap}) have an asymmetry: $|\omega^{(-)}({\bf k})| \neq |\omega^{(+)}({\bf k})|$. In the regime of strong magnetic fields, {\it i. e.}, $|{\bf B}_{0}| \gg |{\bf E}_{0}|$, the previous frequencies can be simplified according to the expressions that follow:
\begin{eqnarray}\label{omegapmB>>E}
\omega^{(\pm)}({\bf k}) \simeq \pm \, |{\bf k}| \, e^{-\gamma} \sqrt{ 1 + 2 \, e^{\gamma} \, \sinh(\gamma) \, \cos^2\theta } \; ,
\end{eqnarray}
where $\cos\theta=\hat{{\bf k}}\cdot\hat{{\bf B}}_{0}$. It is interesting to highlight that this frequency coincides with the results given in ref. \cite{Sorokin1}, where the authors investigated a different configuration of external fields, namely, (anti)parallel external fields. 
In this condition, the frequency depends on the direction of the magnetic field relative to the ${\bf k}$-wave vector, and consequently, the refractive index also depends on the $\theta$-angle. In addition, the frequency (\ref{omegapmB>>E}) does not depend on the magnetic field magnitude when $|{\bf B}_{0}| \rightarrow \infty$. We also point out that the frequencies (\ref{omegam}) and (\ref{omegap}) are degenerate at $\omega^{(+)}=\omega^{(-)}=d_{E}\, {\bf B}_{0}\cdot({\bf k}\times{\bf E}_{0})$ if the ModMax parameter and the EM background field
satisfy the condition
\begin{eqnarray}\label{condd3EB}
1 + d_{E}(\hat{{\bf k}}\cdot{\bf B}_{0})^{2}+d_{E}\,{\bf B}_{0}^2
=d_{E}(\hat{{\bf k}}\times{\bf E}_{0})^{2}+
\nonumber \\
+\,d_{E}^2\left({\bf E}_{0}^2-{\bf B}_{0}^{2} \right)(\hat{{\bf k}}\cdot{\bf B}_{0})^2 \; .
\end{eqnarray}
On the other hand, if the EM background fields satisfy the inequality
\begin{eqnarray}\label{condEB}
1 + d_{E}(\hat{{\bf k}}\cdot{\bf B}_{0})^{2}+d_{E}\,{\bf B}_{0}^2>
d_{E}(\hat{{\bf k}}\times{\bf E}_{0})^{2}+
\nonumber \\
+\,d_{E}^2\left({\bf E}_{0}^2-{\bf B}_{0}^{2} \right)(\hat{{\bf k}}\cdot{\bf B}_{0})^2 \; ,
\end{eqnarray}
both the dispersion relations (\ref{omegam}) and (\ref{omegap}) are real. Otherwise, if the expression (\ref{condEB})
has the opposite inequality $(<)$, the dispersion relations have imaginary parts.
The refractive index associated with the $\omega^{(+)}$-frequency is defined by
\begin{widetext}
\begin{eqnarray}\label{nomegap}
n=
%\frac{|{\bf k}|}{\omega^{(+)}}=
\frac{1+d_{E}\,{\bf B}_{0}^2}{ d_{E} \, {\bf B}_{0}\cdot(\hat{{\bf k}}\times{\bf E}_{0})+\sqrt{1-d_{E}(\hat{{\bf k}}\times{\bf E}_{0})^{2}
-d_{E}^2\left({\bf E}_{0}^2-{\bf B}_{0}^{2} \right)(\hat{{\bf k}}\cdot{\bf B}_{0})^2 + d_{E}\,(\hat{{\bf k}}\cdot{\bf B}_{0})^{2}+d_{E}\,{\bf B}_{0}^2 } } \; ,
\end{eqnarray}
\end{widetext}
where the condition (\ref{condEB}) can be imposed to obtain a real refractive index. It is opportune to point out that eq. (\ref{condEB})
with the opposite inequality $(<)$ implies into a dichroism effect in the refractive index (\ref{nomegap}). In the regime of a
strong magnetic field $(|{\bf B}_{0}|\gg |{\bf E}_{0}|)$, the previous refractive index leads to
\begin{eqnarray}\label{napprox}
n \simeq \frac{ e^{\gamma} }{ \sqrt{1+(e^{2\gamma}-1)\cos^2\theta} } \; ,
%\hspace{0.4cm} \mbox{if} \hspace{0.4cm}
%B_{0} \gg E_{0}  \; .
\end{eqnarray}
which depends on the $\theta$-angle and $\gamma$-parameter.

%Notice that, if the ModMax parameter is $\gamma \gg 1$, the refractive index (\ref{napprox})
%is reduced to $n \simeq |\sec\theta|$. This is the same result obtained in the reference
%\cite{MJNevesPRD2021}, in the case of the Born-Infeld electrodynamics in the presence of a
%strong magnetic background field.
%

%
The group velocity steaming from the polynomial equation (\ref{eqomega4}) is
\begin{widetext}
\begin{eqnarray}\label{vg}
{\bf v}_{g} = \frac{ \omega \, \left[ \, \left(1-d_{E}\,{\bf E}_{0}^2\right)\,{\bf k}+ d_{E}\,({\bf E}_{0}\cdot{\bf k}) \, {\bf E}_{0}
+ d_{E} \, ({\bf B}_{0}\cdot{\bf k}) \, {\bf B}_{0}
+ d_{E} \, \omega \, ({\bf E}_{0}\times{\bf B}_{0}) \, \right] }{ 2\omega^2-{\bf k}^2 + d_{E} \, (\omega\,{\bf B}_0-{\bf k}\times{\bf E}_{0} )^2+d_{E}\,\omega\,{\bf B}_{0}\cdot( \omega\,{\bf B}_0-{\bf k}\times{\bf E}_{0} )-d_{E}\,({\bf B}_{0}\cdot{\bf k})^{2} }  \; ,
\end{eqnarray}
\end{widetext}
where $\omega$ must be evaluated at the solutions $\omega=0$, (\ref{omegam}) and (\ref{omegap}). The external electric and magnetic fields correspond to space anisotropies and the group velocity of the wave is no longer exclusively along {\bf k}; it develops components in other directions too. The limit $\gamma \rightarrow 0$ in eq. (\ref{vg}) recovers the usual result from Maxwell ED,  ${\bf v}_{g}= c \, \hat{{\bf k}}$ (with $c=1$ in natural units). As expected, the solution $\omega=0$
yields a null group velocity. In the regime of a strong magnetic background $(|{\bf B}_{0}| \gg |{\bf E}_{0}|)$, the group velocity (\ref{vg}) is reduced to
\begin{eqnarray}\label{vgapprox}
{\bf v}_{g} \simeq \, e^{-\gamma} \frac{\hat{{\bf k}}+2 \, e^{\gamma} \, \sinh\gamma \, \cos\theta \, \hat{{\bf B}}_{0} }{ \sqrt{ 1+ 2 \, e^{\gamma} \, \sinh(\gamma) \, \cos^2\theta }} \; .
\end{eqnarray}
Observe that this result also depends on the $\theta$-angle %in that the magnetic background vector does with the wave propagation direction. Furthermore,the group velocity
and has a component on the magnetic background direction.
For the sake of simplicity, we consider the vectors ${\bf k}$, ${\bf B}_{0}$ and ${\bf E}_{0}$ perpendicular among themselves, {\it i.e.},
${\bf k}\cdot{\bf B}_{0}={\bf k} \cdot {\bf E}_{0}={\bf B}_{0}\cdot{\bf E}_{0}=0$.
%In this particular case,
%the condition (\ref{condd3EB}) for the degenerated frequencies implies that $\gamma\gg 1$, if $|{\bf E}_{0}|>|{\bf B}_{0}|$.
The refraction index of this medium associated with the $\omega^{(+)}$-frequency is given by
%
%
%\begin{subequations}
\begin{equation}
%n^{(-)} \!&=&\! \frac{1+d_{E}B_{0}^2}{d_{E}B_{0}E_{0}-\sqrt{1-d_{E}\left(E_{0}^2-B_{0}^2\right)}}? \; ,
%\label{np} \\
n_{1} = \frac{E_{0}^2-B_{0}^2+2 \, e^{-\gamma} \sinh(\gamma) \, B_{0}^2}{\left(E_{0}^2-B_{0}^2\right)e^{-\gamma}+2\,e^{-\gamma}\sinh(\gamma)\,E_{0}\,B_{0}} \; ,
\label{n1}
\end{equation}
%\end{subequations}
%
when $E_{0}>B_{0}$, and
\begin{equation}
%n^{(-)} = \frac{1+d_{E}B_{0}^2}{d_{E}B_{0}E_{0}-\sqrt{1-d_{E}\left(E_{0}^2-B_{0}^2\right)}}? \; ,
%\label{np} \\
n_{2} = \frac{E_{0}^2-B_{0}^2-2 \, e^{\gamma} \, \sinh(\gamma) \, B_{0}^2}{\left(E_{0}^2-B_{0}^2\right)e^{\gamma}-2\,e^{\gamma}\sinh(\gamma)\,E_{0}\,B_{0}} \; ,
\label{n2}
\end{equation}
for the condition $B_{0}>E_{0}$. Here, we denote $E_{0}=|{\bf E}_{0}|$ and $B_{0}=|{\bf B}_{0}|$.
The $\gamma \rightarrow 0$ limit leads to $n_1=n_2=1$. Although the $d_{E}$ coefficient is not defined for
$E_{0} \rightarrow B_{0}$, this limit leads to $n_1=n_2=1$, if $E_{0} \neq B_{0}$. For the condition of a strong magnetic field,
$B_{0} \gg E_{0}$, the refraction index $n_{2} \simeq e^{-\gamma}$ does not depend on the magnetic field.
%
%The refractive index (\ref{n1}) and (\ref{n2}) are plotted in the fig. (\ref{n12x}) as function of the dimensionless variable $x:=B_{0}/E_{0}$ for the %values $\gamma=0.5$ (black line), $\gamma=1.0$ (blue line) and $\gamma=1.5$ (red line), respectively. The curves meet at
%$E_{0}=B_{0} \, (x=1)$, that correspond to the refractive index at $n_1=n_2=1$.
%
%
%
%\begin{figure*}[th]
%\vspace{-5pt}
%\centering
%\includegraphics[width=0.45\textwidth]{n1x.eps} \quad\quad
%\includegraphics[width=0.45\textwidth]{n2x.eps}
%\caption{The refractive index as function of $x=B_{0}/E_{0}$ for the $\gamma$-values $\gamma=0.5$ (black line), $\gamma=1.0$ (blue line)
%and $\gamma=1.5$ (red line).
%Left panel : The refractive index (\ref{n1}) as function of $x$. Right panel : The refractive index (\ref{n2}) as function of $x$.}
%\label{n12x}
%\end{figure*}
%

%

%
%If we multiply the solutions (\ref{omegaBpm}) by $\hbar$, and
%From the De Broglie duality correspondence, the energy-momentum relations for the propagating excitations read as below:
%
%\begin{subequations}
%\begin{eqnarray}
%E_{1}^2 \!&=&\! {\bf p}^2 \left[ \, 1-\frac{d_1 }{c_1} \, ({\bf B}\times \hat{{\bf p}})^2 \, \right] \; ,
%\label{E1} \\
%E_{2}^2 \!&=&\! {\bf p}^2 \left[ \, 1-\frac{d_2 \, ({\bf B}\times \hat{{\bf p}})^2 }{ c_1+d_{2}\,{\bf B}^2 } \, \right] \; .
%\label{E2}
%\end{eqnarray}
%\end{subequations}
%

%Maxwell case is recovered when $c_1=1$ and $d_{1}=d_{2}=0$ in (\ref{eqomega4}).
%The result (\ref{omega2B}) is recovered by taking $d_{2}=d_{3}=0$.
%%
%

%
%
Under the conditions of ${\bf k}\cdot{\bf B}_{0}={\bf k} \cdot {\bf E}_{0}={\bf B}_{0}\cdot{\bf E}_{0}=0$, the group velocity associated with the previous frequencies is obtained from eq. (\ref{vg}), namely,
\begin{equation}\label{vgkEB}
{\bf v}_{g}= \frac{ \hat{{\bf k}} \, \omega \, \left[ \, k-d_{E}\,B_{0}\,E_{0}\,\omega-d_{E}\,E_{0}^2\,k \, \right] }{2\omega^2-k^2+d_E(\omega B_{0}-k\,E_{0})^2+d_{E}\,\omega\,B_{0} (\omega B_{0} -kE_{0}) } \; .
\end{equation}
%
%where $\omega$ must be evaluated at (\ref{omegap}), and substituting the $\omega^{(+)}$ frequency,
%we obtain :
%
%
%\begin{subequations}
%\begin{eqnarray}
%v_{1g}^{(+)} \!\!&=&\!\! \frac{ E_{0}+e^{\gamma}\,B_{0} }{e^{\gamma} \, E_{0}+B_{0} }
%\;\; \mbox{if} \;\; E_{0} > B_{0} \; , \;\;
%\label{vgp1} \\
%v_{2g}^{(+)} \!\!&=&\!\! e^{\gamma} \, \frac{ B_{0}^2-E_{0}^2+2E_{0} \, B_{0} \,\sinh(\gamma) }{B_{0}^2\,e^{2\gamma}-E_{0}^2}
%\;\; \mbox{if} \;\; B_{0} > E_{0} \; . \;\;\;
%\hspace{0.7cm}
%\label{vgp2}
%\end{eqnarray}
%\end{subequations}
%
%The limit $\gamma \rightarrow 0$ recovers the usual result : $\lim_{\gamma \rightarrow 0}v_{1g}^{(+)}=\lim_{\gamma \rightarrow 0}v_{2g}^{(+)}=1$ ($c=1$ in %natural units).
%
For a strong magnetic field, we have ${\bf v}_{g} \simeq \, e^{-\gamma} \, \hat{{\bf k}} $,
the group velocity decays with the ModMax parameter and propagates only on the $\hat{{\bf k}}$-direction. 
This confirms the result (\ref{vgapprox}) whenever $\theta=\pi/2$. \\

\section{Birefringence in presence of electric and magnetic backgrounds}
\label{sec5}

In this Section, we investigate the birefringence phenomenon in the context of ModMax ED. For this purpose, we need to obtain the variation of the refractive index in relation to the magnetic background field. In what follows, we assume the external fields as ${\bf B}_{0}=B_{0} \, \hat{{\bf z}}$ and ${\bf E}_{0}=E_{0} \, \hat{{\bf y}}$.

%The phenomenon of birefringence analysis in the ModMax ED starts in an electromagnetic background field in which we assume the variation of the refractive index in relation to the magnetic background field. We consider the external magnetic fieldon the $z$-direction and the electric field on the $y$-direction,{\it i. e.}, ${\bf B}_{0}=B_{0} \, \hat{{\bf z}}$ and ${\bf E}_{0}=E_{0} \, \hat{{\bf y}}$, respectively.

Initially, let us consider the plane wave solution for the electric field
into the form ${\bf e}(x,t)=e_{03}\,\hat{{\bf z}} \, e^{i \left(k x-\omega t\right)}$. Notice that
the propagation direction is ${\bf k}=k\,\hat{{\bf x}}$ and the wave amplitude is parallel to the magnetic background field. 
Under these conditions, the wave equation (\ref{EqMatrix}) yields the parallel refractive index :

\begin{eqnarray}\label{nparallel}
n_{\parallel}=\sqrt{ \mu_{22}(E_{0},B_{0})\,\varepsilon_{33}(E_{0},B_{0}) }
=\sqrt{ \frac{1+d_{E}\,B_{0}^2}{1-d_{E}\,E_{0}^2} } \; .
\end{eqnarray}

\begin{figure*}[th]
%\vspace{-5pt}
\centering
\includegraphics[width=0.45\textwidth]{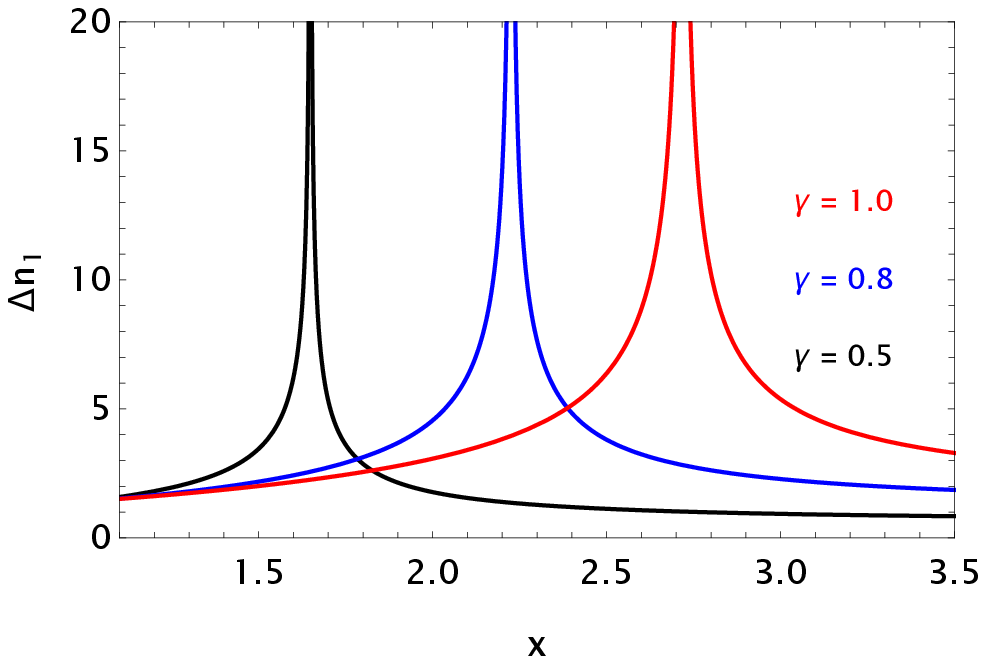} \quad\quad
\includegraphics[width=0.44\textwidth]{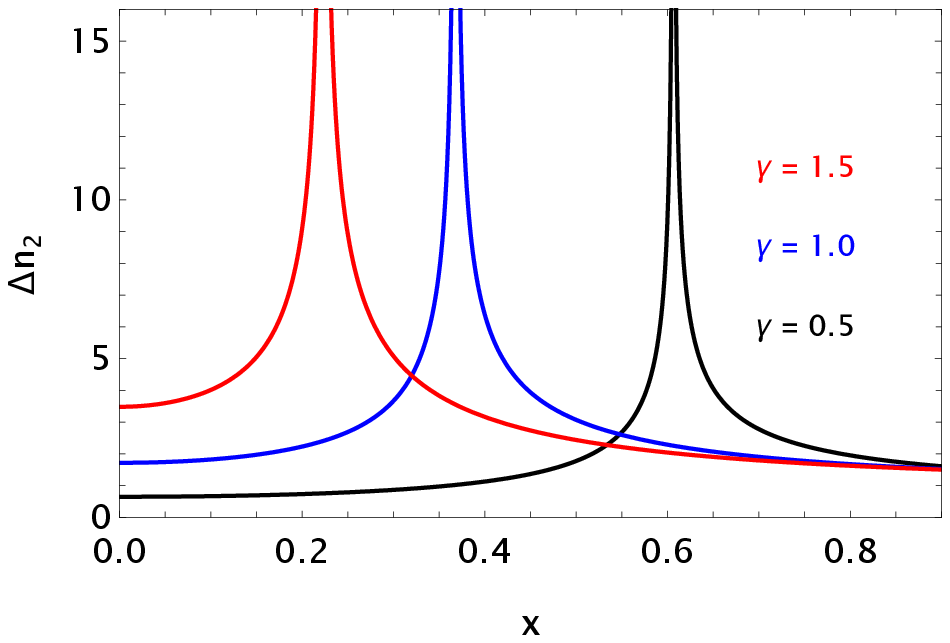}
\caption{The variation of the refractive index $\Delta n_{1}$ (left panel) and $\Delta n_{2}$ (right panel) as functions of the dimensionless variable $x=E_{0}/B_{0}$. The left panel is plotted for $\gamma=0.5$ (black line), $\gamma=0.8$ (blue line) and $\gamma=1.0$ (red line) in the range of $E_{0}>B_{0}$. The right panel is for $\gamma=0.5$ (black line), $\gamma=1.0$ (blue line) and $\gamma=1.5$ (red line), with the condition of $B_{0}>E_{0}$, respectively. }
\label{fig2}
\end{figure*}

In the second situation, the plane wave solution has the amplitude
perpendicular to the magnetic background field,
${\bf e}(x,t)=e_{02}\,\hat{{\bf y}} \, e^{i \left(k x-\omega t\right)}$. In this case,
the perpendicular refractive index is
\begin{eqnarray}\label{nperpendicular}
n_{\perp}=\sqrt{ \mu_{33}(E_{0},B_{0}) \, \varepsilon_{22}(E_{0},B_{0}) } = 1 \; .
\end{eqnarray}
The birefringence is defined by the difference of the refractive indices :
\begin{eqnarray}\label{Deltan}
\Delta n=|n_{\parallel}-n_{\perp}|=\left| \, \sqrt{ \frac{1+d_{E}\,B_{0}^2}{1-d_{E}\,E_{0}^2} }-1 \, \right| \; .
\end{eqnarray}
%
%in which $\Delta n>0$, if $\gamma > 0$.
In the limit $\gamma \rightarrow 0$, we have $\Delta n=0$ and birefringence disappears.
Using the definitions of $d_{E}$ in eqs. (\ref{dEEB}) and (\ref{dEBE}), we obtain the following difference
\begin{subequations}
\begin{eqnarray}
\Delta n_{1} \!&=&\! \left| \, \sqrt{ \frac{ E_{0}^2 \, e^{2 \gamma } - B_{0}^2 }{ E_{0}^2 -B_{0}^2 \, e^{2 \gamma } } }-1 \, \right|
\; , \;\; \mbox{if} \;\; E_{0} > B_{0} \; , \;
\label{Deltan1}
\\
\Delta n_{2} \!&=&\! \left| \, \sqrt{\frac{B_{0}^2 \, e^{2 \gamma }-E_{0}^{2} }{B_{0}^2 -E_{0}^2 \, e^{2 \gamma } }}-1 \, \right|
\; , \;\; \mbox{if} \;\; B_{0} > E_{0} \; . \;
\label{Deltan2}
\end{eqnarray}
\end{subequations}
The plot of $\Delta n_{1}$ versus $x \, (x=E_{0}/B_{0})$ (left panel), and $\Delta n_{2}$ versus $x$ (right panel)
are illustrated in fig. (\ref{fig2}). The vertical asymptote in these plots is located at $E_{0}= e^{\gamma} \, B_{0}$ (left panel), and at $E_{0}= e^{-\gamma} \, B_{0}$ in the right panel. Imposing the causality and unitarity conditions on the ModMax ED \cite{Sorokin1}, we choose the $\gamma$-values with $\gamma>0$. In the plot of $\Delta n_{1}$, we choose the values for the $\gamma$-ModMax parameter :
$\gamma=0.5$ (black line), $\gamma=0.8$ (blue line) and $\gamma=1.0$ (red line), in which $E_{0}>B_{0}$. The $\gamma$-values in the
plot of $\Delta n_{2}$ are $\gamma=0.5$ (black line), $\gamma=1.0$ (blue line) and $\gamma=1.5$ (red line), that satisfies the condition
$B_{0}>E_{0}$.
Birefringence manifests in the region below the curves in fig. (\ref{fig2}).
Under an intense magnetic background field, $B_{0} \gg E_{0}$ (or when $E_{0} \rightarrow 0$) and $\gamma \ll 1$, the second solution (\ref{Deltan2}) reads as given below:
\begin{eqnarray}
\Delta n_{2}\simeq \gamma \; ,
\end{eqnarray}
which agrees with the result given in ref. \cite{Sorokin2}. Similarly, if the electric background field is intense, where $E_{0} \gg B_{0}$ (or when $B_{0} \rightarrow 0$) and $\gamma \ll 1$, the first solution in eq. (\ref{Deltan1}) leads to the same result $\Delta n_{1}\simeq \gamma$. The birefringence curves go to infinity when $x \rightarrow e^{\gamma}$ (left panel) and $x \rightarrow e^{-\gamma}$ (right panel). \\
\section{Interaction energy for an Axionic ModMax electrodynamics under a uniform magnetic field}
\label{sec6}
The coupling of the axion to the photon has been widely investigated in different scenarios. There is a rich literature on the issue, with special attention to the Primakoff effect (see, for instance, the review \cite{Sikivie} and references therein). In the present Section, we pursue the investigation of the low-energy interparticle potential that emerges from a scenario where the axion is coupled to the photon as described by our object of study, ModMax ED. It is true that we are taking an unusual viewpoint, by bringing together two new physics: ModMax with its $\gamma$-parameter, and the axionic sector, with the axion mass, $m_{a}$, and the axion-photon coupling constant, $g_{a\gamma\gamma}$. We justify our attitude by calling into question our interest in understanding how the mentioned three parameters interfere with each other in the screening and the confining sectors of the particle-antiparticle potential, after we integrate out the axion contributions to get an effective photonic mediation of the interaction. Moreover, if an external magnetic field is switched on, it would be interesting to understand how ${\gamma}$, $m_{a}$, $g_{a\gamma\gamma}$ and the external magnetic field, $\bf B$ , combine to screen and to  a confining tail to the interparticle potential. This shall be the content of this Section.

As already expressed, we shall now discuss the interaction energy between static point-like sources for an axionic ModMax electrodynamics under a uniform magnetic field, along the lines of refs. \cite{Gaete97,GEN_BI,LOG,Gaete_AHEP_2021,Gaete_EPJC_2022}. This can be done by computing the expectation value of the energy operator $H$ in the physical state $\left| \Phi  \right\rangle$, which we denote by $\langle H\rangle _\Phi$. However, before going to the derivation of the interaction energy, we will describe very briefly the model under consideration. The initial point of our analysis is the Lagrangian density
\begin{eqnarray}
{\cal L} &=& \cosh \left( \gamma  \right){\cal F} + \sinh \left( \gamma  \right)\sqrt {{{\cal F}^2} + {{\cal G}^2}}  + \frac{1}{2}{\left( {{\partial _\mu } a} \right)^2}
\nonumber \\
&&
\hspace{-0.5cm}
-\frac{1}{2} \, {m_{a}^2} \, a^2
+ {g_{a\gamma \gamma }}\, a\, {\cal G} \; , \label{axion05}
\end{eqnarray}
where $a$ is the axion field and $g_{a\gamma \gamma }$ has dimension of ${(\mbox{mass})^{ - 1}}$.

%$m_{a}$ is the axion mass, ${g_{a\gamma \gamma }}$ is the axion-photon coupling (with dimension ${(\mbox{mass})^{ - 1}}$).
%

%
Following our earlier procedure \cite{Gaete_AHEP_2021,Gaete_EPJC_2022}, if we consider the model in the limit of a very heavy $a$-field and we are bound to energies much below $m_{a}$, we are allowed to integrate over $a$ and to speak about an effective model for the $A^{\mu}$-ModMax field. Once this is done, we arrive at the following effective Lagrangian density :
\begin{eqnarray}
{\cal L} &=& \cosh \left( \gamma  \right){\cal F} + \sinh \left( \gamma  \right)\sqrt {{{\cal F}^2} + {{\cal G}^2}}
\nonumber \\
&+& \frac{{{g_{a\gamma \gamma }^2}}}{2}\,{\cal G} \, \frac{1}{{\left( {\Box  + {m_{a}^2}} \right)}} \, {\cal G} \; , \label{axion10}
\end{eqnarray}
where we define the D'Alembertian operator $\Box  \equiv {\partial_\mu }{\partial^\mu }$.
Since we are interested in estimating the lowest-order interaction energy, we will linearize the above effective theory following the procedure that led to the equation (\ref{LMM2}). Thus, in the case of a pure magnetic background, we make ${\bf E}={\bf 0}$, in which the effective Lagrangian density simplifies to
\begin{eqnarray}
{\cal L} \!&=&\!  - \frac{1}{4} \, {e^{ - \gamma }} \, {f_{\mu \nu }} \, {f^{\mu \nu }}
- \frac{1}{2} \, {e^{ - \gamma }} \, F_{B\mu \nu }\,{f^{\mu \nu }}
\nonumber \\
&&
\hspace{-0.5cm}
+ \frac{1}{4} \, \frac{{\sinh \left( \gamma  \right)}}{{{\bf B}_0^2}} \, \tilde{F}_{B\mu \nu }\tilde{F}_{B\kappa \lambda } \, {f^{\mu \nu }} \, {f^{\kappa \lambda }}
\nonumber \\
&&
\hspace{-0.5cm}
+ \frac{{{g_{a\gamma \gamma }^2}}}{8} \, \tilde F_{B\mu \nu } \, {f^{\mu \nu }} \, \frac{1}{{\left( {\Box  + {m_{a}^2}} \right)}}
\, \tilde F_{B\kappa \lambda }\,{f^{\kappa \lambda }} \; .
\label{axion15}
\end{eqnarray}
In passing, we note that the effective model described by the Lagrangian density (\ref{axion15}) is a theory with non-local time derivative. However, as we have already explained in refs. \cite{Gaete_AHEP_2021,Gaete_EPJC_2022}, we recall that this section is aimed at studying the static potential, so that $\Box$ can be replaced by the spatial operator $-{\nabla ^2}$. As before, we will maintain $\Box$, but it should be borne in mind that this paper essentially deals with the static case.
With this in hand, the canonical quantization of this effective theory from the Hamiltonian analysis point of view is straightforward. The canonical momenta reads
\begin{eqnarray}
{\Pi ^\mu } \!&=&\!  - {e^{ - \gamma }}{f^{0\mu }} - {e^{ - \gamma }} \, F_{B}^{\;\;\,0\mu } + \frac{{\sinh \left( \gamma  \right)}}{{{\bf B}_0^2}} \, \tilde F_{B\kappa \rho }\,{f^{\kappa \rho }} \, \tilde F_{B}^{\;\;\,0\mu }
\nonumber \\
&&
\hspace{-0.5cm}
+\frac{{{g_{a\gamma \gamma }^2}}}{2} \, \tilde F_{B\kappa \rho }{f^{\kappa \rho }}
\,\frac{1}{{\left( {\Box  + {m_{a}^2}} \right)}} \, {\tilde F_{B}^{\;\;\,0\mu }} \, , \label{axion20}
\end{eqnarray}
which produces the usual primary constraint
\begin{equation}\label{axion25-a}
\Pi^0 = 0 \; ,
\end{equation}
and
\begin{equation}
\Pi_{i} =  - \left\{ {e^{ - \gamma }}\, {{\delta _{ij}} + 2{B_i}{B_j} \left[ {\frac{{\sinh \left( \gamma  \right)}}{{{\bf B}_0^2}} + \frac{{{\raise0.5ex\hbox{$\scriptstyle {{g_{a\gamma \gamma }^2}}$}
\kern-0.1em/\kern-0.15em
\lower0.25ex\hbox{$\scriptstyle 2$}}}}{{\left( {\Box  + {m_{a}^2}} \right)}}} \right]} \right\}{e_j} \; .
\label{axion25-b}
\end{equation}

Let us also mention here that the electric field due to the fluctuation takes the form
\begin{equation}\label{axion30}
{e_i} = \frac{1}{{u\det D}}\left( {{\delta _{ij}}\det D - \frac{1}{{{\Omega ^2}}}{B_i}{B_j}} \right){\Pi _j} \; ,
\end{equation}
where $u = {e^{ - \gamma }}$ and $\det D = 1 + \frac{{{{\bf B}^2}}}{{{\Omega ^2}}}$, whereas
\begin{eqnarray}
\frac{1}{{{\Omega ^2}}} = 2 \, {e^\gamma }\left[ \, {\frac{{\sinh \left( \gamma  \right)}}{{{\bf B}^2}} + \frac{{{\raise0.5ex\hbox{$\scriptstyle {{g_{a\gamma \gamma }^2}}$}
\kern-0.1em/\kern-0.15em
\lower0.25ex\hbox{$\scriptstyle 2$}}}}{{\left( {\Box  + {m_{a}^2}} \right)}}} \, \right] \; .
\end{eqnarray}
Recalling again that ${\bf B}$ represents the external (background) magnetic field around which the $a^{\mu}$-field fluctuates.
The canonical Hamiltonian can be worked as usual and is given by
\begin{eqnarray}
{H_C} \!&=&\! \int {{d^3}x \left[ \, {{\Pi ^i} \, {\partial _i}{a_0} + \frac{{{{\bf \Pi} ^2}}}{{2u}} + \frac{1}{2}u\,{{\bf b}^2} + u\,{\bf B} \cdot {\bf b}} \,\right]}
\nonumber \\
&&
\hspace{-0.5cm}
-\int {{d^3}x} \;\,  \frac{ 1 }{{2u\,{\Omega ^2}\,\det D}} \, \left( {{\bf B} \cdot {\bf b}} \right)^2 \, . \label{axion35}
\end{eqnarray}

Time conserving the primary constraint, ${\Pi}_0$, immediately yields the secondary constraint, ${\Gamma _1} \equiv {\partial _i}{\Pi ^i} = 0$, which is the Gauss constraint, and together displays the first-class structure of the theory. The extended Hamiltonian that generates translations in time is now found to be
\begin{eqnarray}
H = H_C + \int {d^3 } x\left[ \, {c_0 \left( x \right) \Pi _0 \left( x \right) + c_1
\left( x\right)\Gamma _1 \left( x \right)} \, \right] \; ,
\end{eqnarray}
where $c_0 \left( x\right)$ and $c_1 \left( x \right)$ are the Lagrange multipliers. As before, neither $a^0 $ nor $\Pi^0$ are of interest in describing the system and may be discarded of the theory. Thus we are left with the following expression for the Hamiltonian
\begin{equation}
H = \int {{d^3}x\left[ \, {c\left( x \right){\partial _i}{\Pi ^i} + \frac{{{{\bf \Pi} ^2}}}{{2\,u}} - \frac{1}{{2\,u\,{\Omega ^2}\det D}}
\,{{\left( {{\bf B} \cdot {\bf b}} \right)}^2}} \, \right]} \, , \label{axion40}
\end{equation}
we have defined $c(x) = c_1 (x) - a_0 (x)$.
To fix gauge symmetry we adopt the gauge discussed previously \cite{Gaete97}, that is,
\begin{equation}
\Gamma _2 \left( x \right) \equiv \int\limits_{C_{\zeta x} } {dz^\nu }
\, a_\nu\left( z \right) \equiv \int\limits_0^1 {d\lambda \, x^i } a_i \left( {
\lambda x } \right) = 0 \, . \label{axion45}
\end{equation}
Here $\lambda$ $(0\leq \lambda\leq1)$ is the parameter describing the
space-like straight path $x^i = \zeta ^i + \lambda \left( {x - \zeta}
\right)^i $, and $\zeta^{i}$ is a fixed point (reference point).
There is no essential loss of generality if we restrict our considerations to $\zeta^i=0$.
With such a choice, the fundamental Dirac bracket is given by
\begin{eqnarray}
\left\{ {a_i \left( {\bf x} \right),\Pi ^j \left( {\bf y} \right)} \right\}^{\ast} \!&=&\! \delta_{i}^{\;\,j} \, \delta ^{\left( 3 \right)} \left( {{\bf x} - {\bf y}} \right)
\nonumber \\
&&
\hspace{-0.5cm}
-\partial_i^x
\int\limits_0^1 {d\lambda \, x^j } \delta ^{\left( 3 \right)} \left( {\lambda
{\bf x}- {\bf y}} \right) \, . \label{axion50}
\end{eqnarray}
Next, we recall that the physical states $\left| \Phi \right\rangle $ are gauge-invariant \cite{Gaete_AHEP_2021,Gaete_EPJC_2022}. In that case we consider the stringy gauge-invariant state
\begin{eqnarray}
\left| \Phi  \right\rangle  \!&\equiv&\! \left|\, {\overline{\Psi} \left( {\bf y} \right)\Psi \left( {{{\bf y}^ {\prime} }} \right)} \,\right\rangle  \nonumber\\
&=& \overline{\Psi} \left( {\bf y} \right)\exp \left( {iq\int_{{{\bf y}^ {\prime} }}^{\bf y} {d{z^i}{a_i}\left( z \right)} } \right)\Psi \left( {{{\bf y}^ {\prime} }} \right)\left| 0 \right\rangle \, , \label{axion55}
\end{eqnarray}
where the line integral is along a space-like path on a fixed time slice, $q$ is the fermion charge and $\left| 0 \right\rangle$ is the physical vacuum state.

This leads us to the expectation value ${\left\langle H \right\rangle _\Phi }$
\begin{equation}
{\left\langle H \right\rangle _\Phi } = {\left\langle H \right\rangle _0} + \left\langle H \right\rangle _\Phi ^{\left( 1 \right)} \; , \label{axion60}
\end{equation}
where ${\left\langle H \right\rangle _0} = \left\langle 0 \right|H\left| 0 \right\rangle$, whereas the $\left\langle H \right\rangle _0^{\left( 1 \right)}$ term is given by
\begin{equation}
\left\langle H \right\rangle _\Phi ^{\left( 1 \right)} =  - \frac{{{e^{ - \gamma }}}}{2}\left\langle \Phi  \right|\int {{d^3}x} \, {\Pi ^i} \, \frac{{\left( {{\nabla ^2} - {m_{a}^2}} \right)}}{{\left( {{\nabla ^2} - {M^2}} \right)}} \, {\Pi _i} \, \left| \Phi  \right\rangle \; , \label{axion65}
\end{equation}
where ${M^2} = {m_{a}^2} + {g_{a\gamma \gamma }^2} \, {{\bf B}^2} \, {e^{ - \gamma }}$.

Following our earlier procedure \cite{Gaete_AHEP_2021,Gaete_EPJC_2022}, when $g_{a\gamma \gamma } \to 0$, the static potential profile for two opposite charges located at ${\bf y}$ and ${\bf y}^{\prime}$ then reads
\begin{equation}
V(L) =  - \frac{{{q^2}}}{{4\pi }} \, {e^{ -\gamma }} \, \frac{1}{L} \, , \label{axion70a}
\end{equation}
where $L \equiv |{\bf y} - {{\bf y}^ {\prime} }|$ is the distance that separates the two charges. It is also, up to the ${e^{ - \gamma }}$ factor, just the Coulomb potential. This result agrees with that of ref. \cite{Sorokin3}, and finds here an independent derivation. While in the case $g_{a\gamma \gamma } \ne 0$, the interaction energy takes the form
\begin{equation}
V(L) =  - \frac{{{q^2}}}{{4\pi }} \, {e^{ - \gamma }} \, \frac{{{e^{ - ML}}}}{L} + \frac{{{q^2}{m_{a}^2}}}{{8\pi }}\,{e^{ - \gamma }}\left[ {\ln \left( {1 + \frac{{{\Lambda ^2}}}{{{M^2}}}} \right)} \right]L \, , \label{axion70b}
\end{equation}
where $\Lambda$ is a cutoff. The next step is to give a physical meaning to the cutoff $\Lambda$. Proceeding in the same way as we did before \cite{Gaete_AHEP_2021,Gaete_EPJC_2022}, we recall that our effective model for the electromagnetic field is an effective theory that arises from the integration over the $a$-field, whose excitation is massive. In this case, the Compton wavelength of this excitation ($\ell=m_{a}^{-1}$) defines a correlation distance. In view of this situation, we see that physics at distances of the order or lower than $m_{a}^{-1}$ must necessarily take into account a microscopic description of the $a$-fields. By this we mean that, if we work with energies of the order or higher than $m_{a}$, our effective description with the integrated effects of $a$ is no longer sensible. As a consequence of this, we can identify $\Lambda$ with $m_{a}$. This then implies that the static potential profile assumes the form
\begin{equation}
V(L) =  - \frac{{{q^2}}}{{4\pi }}\,{e^{ - \gamma }}\,\frac{{{e^{ - ML}}}}{L} + \frac{{{q^2}{m_{a}^2}}}{{8\pi }}\,{e^{ - \gamma }}\left[ {\ln \left( {1 + \frac{{{m_{a}^2}}}{{{M^2}}}} \right)} \right]L \, .  \label{axion75}
\end{equation}

Again, up to the ${e^{ - \gamma }}$ factor, we mention that similar forms of interaction potentials have been reported before from different viewpoints. For example, in the context of the Standard Model with an anomalous triple gauge boson couplings \cite{Gaete_AHEP_2021}, in connection with anomalous photon and $Z$-boson self couplings from the Born-Infeld weak hypercharge action \cite{Gaete_EPJC_2022}, also in a theory of antisymmetric tensor fields that results from the condensation of topological defects \cite{Gaete_PLB_2004}, and in a Higgs-like model \cite{Gaete_PLB_2009}.

%
%%%%%%%%%%%%%%%%%%%%%%%%%%%%%%%%%%%%%%%%%%%%%%%%%%%%%%%%%%%%%%%%%%%%%%%%%%%%%%%%%%%%%%%%%%%%%%%%%%%%%%%%%%%%%%%%%%%%%%%%%%%%%%%%%%%%%%%%%%%%%%%%%%%%%%%%%%%%%%%%%%%%%%%%%%%%%%%%%%%%%%%%%%%%%%%%%%%%%%%%%%%%%%%%%%%%%%%%%%%%%%%%%%%%%%%%%%%%%%%%%%%%%%%%%%%%%%%%%%%%%%%%%%%%%%%%%%%%%%%%%%%%%%%%%%%%%%%%%%%%%%%%%%%%%%%%%%%%%%%%%%%%%%%%%%%%%%%%%%%%%%%%%%%%%%%%%%%%%%%%%%%%%%%%%%%%%%%%%%%%%%%%%%%%%%%%%%%%%%%%%%%%%%%%%%%%%%%%%%%%%%%%%%%%%%%%%%%%%%%%%%%%%%%%%%%%%%%%%%%%%%%%%%%%%%%%%

\section{Conclusions}
\label{sec7}
%

%In this contribution, we study the modified Maxwell electrodynamics(ModMax ED) and their propagation properties in a uniform electromagnetic (EM) background field.The ModMax lagrangian is expanded up to second order in the propagating fields around {\color{red} a} uniform EM field. We obtain the permittivity and permeability tensors, the properties of the wave propagation in the presence of a general EM background. We discuss the results of the dispersion relations, refractive index and the group velocity as functions of the wave propagation direction, of the external electric and magnetic. In the regime of a strong magnetic field, the dispersion relations for plane wave solutionsdo not depend on the electric and magnetic field magnitude, and it depends only on the ModMax parameterand on the direction in which the magnetic field does with the wave propagation direction.

%
In this contribution, we study the modified Maxwell electrodynamics
(ModMax ED) and its propagation properties in uniform electric and magnetic background fields.
By expanding the ModMax Lagrangian density up to second order in the propagating photon field, we
obtain the permittivity and permeability tensors, dispersion relations, refractive index and group velocity as functions of the wave propagation direction, ModMax parameter and external fields. In the regime of a strong magnetic field, the dispersion relations for plane wave solutions depend only on the ModMax parameter and  $\theta$-angle between the wave propagation direction and magnetic background field. In this regime, we also find that the refractive index and group velocity decay with the ModMax parameter in the situation where the wave vector, the electric and magnetic background fields are perpendicular to each other.

%The particular case in which the wave vector, the electric, and the magnetic backgrounds are perpendicular among themselves, the refractive index and the group velocity of the medium decays with the ModMax parameter in a regime of strong magnetic field.

We discuss the birefringence phenomenon by taking into account the difference of the refractive indices for the wave amplitude parallel and perpendicular to the external magnetic field. We obtain the region of birefringence in the plots of fig. (\ref{fig2}). In both regimes of strong magnetic or electric fields, birefringence is approximately given by the ModMax parameter, which confirms the result obtained previously in the ref. \cite{Sorokin2}. The well-known results of Maxwell ED are recovered in the limit where the ModMax parameter goes to zero.
Finally, using the gauge-invariant but path-dependent variables formalism, we have computed the static potential profile for an axionic ModMax electrodynamics under a uniform magnetic field. Once again, we have exploited a correct identification of field degrees of  freedom with observable quantities. Interestingly, the static potential profile contains a linear potential  leading to the confinement of static charges. As already expressed similar forms of interaction potentials have been reported before from different viewpoints \cite{Gaete_AHEP_2021,Gaete_EPJC_2022,Gaete_PLB_2004,Gaete_PLB_2009}.
Having in mind the possible relevance of the ModMax model
to describe non-linear electromagnetic effects, we call into
question its application to study a number of physical properties
of Dirac materials. In the work of ref. \cite{CondMatter}, the authors show that
the latter may display electromagnetic non-linearities at magnetic
fields as low as $1T$. We point out that reassessing the inspection
of magnetic enhancement of the dielectric constant of insulators
and, on the other hand, possible electric modulation of magnetization
could be a good path to further investigate the potentialities of ModMax.
A direct contact we might establish between ModMax and electroweak
physics could be through the issue of the photon and $Z$-boson
self-couplings by associating the weak hypercharge symmetry to a
ModMax description. In the work of Ref. \cite{Gaete_EPJC_2022},
we adopt a Born-Infeld description for the weak hypercharge and
consider the $Z$-decay channel into three photons to constrain the Born-
Infeld parameter. By going along the same lines with ModMax, we could
get a bound on the $\gamma$-parameter by considering the $Z$-three photon
anomalous vertex and the decay of the $Z$-boson into three photons.

\section*{Acknowledgments}

L. P. R. Ospedal is grateful to the CNPq - Conselho
Nacional de Desenvolvimento Científico e Tecnológico
(Brazil) - supported by Grant 166386/2020-0 for his postdoctoral fellowship. P. Gaete was partially supported by
Grant No. ANID PIA/APOYO AFB220004 (Chile).

%\ni
%The authors express their gratitude to P. Gaete and A. Spallicci for stimulating discussions on diverse aspects of non-linear electrodynamic models.
%{\color{red} F. Karbstein and D. Sorokin are also deeply acknowledged for pointing out important references and drawing our attention to relevant aspects of the Euler-Heisenberg and Born-Infeld Electrodynamics, respectively.} M. J. Neves thanks CNPq (Conselho Nacional de Desenvolvimento Cient\' ifico e Tecnol\'ogico), Brazilian scientific support federal agency, for partial financial support, Grant number 313467/2018-8. L.P.R. Ospedal is grateful to the Ministry for Science, Technology and Innovations (MCTI) and CNPq for his Post-Doctoral Fellowship under the Institutional Qualification Program (PCI).

%
%\section{References}
%


\begin{thebibliography}{30}

\bibitem{BornI} M. Born, L. Infeld, {\it Foundations of the new field theory}, Proc. R. Soc. Lond. Ser. A {\bf 144}, 425 (1934).

\bibitem{Fermi1} E. Fermi, {\it Sopra l'Elettrodinamica Quantistica}, Rend. Accad. Lincei {\bf 9} (1929) 881.

\bibitem{Fock} P. A. M. Dirac, V. A. Fock and B. Podolsky, {\it On quantum Electrodynamics}, Physikalische Zeitschrift der Sowjetunion, Band 2, Heft 6 (1932).

\bibitem{Fermi2} E. Fermi, {\it Quantum theory of radiation}, Rev. Mod. Phys. {\bf 4} (1932) 87.

\bibitem{Dirac} P. A. M. Dirac, {\it On quantum electrodynamics}, Physikalische Zeitschrift der Sowietunion 2 (1933) 468.

\bibitem{HEuler} H. Euler and W. Heisenberg, {\it Consequences of Dirac Theory of the Positron}, Z. Phys. {\bf 98}, 714 (1936) [arXiv:physics/0605038].

\bibitem{Adler} S.L. Adler, {\it Photon splitting and photon dispersion in a strong magnetic field}, Ann. Phys. (N.Y.) {\bf 67}, 599 (1971).

\bibitem{Pistoni} V. Constantini, B. De Tollis, G. Pistoni, {\it Nonlinear effects in quantum electrodynamics}, Nuovo Cimento A {\bf 2}, 733 (1971).

\bibitem{Ruffini} R. Ruffini, G. Vereshchagin and S-S. Xue,  {\it Electron-positron pairs in physics and astrophysics: from heavy nuclei to black holes}, Phys. Rept. {\bf 487},  1 (2010).

\bibitem{Dunne} G. V. Dunne,  {\it The Euler-Heisenberg effective action: 75 years on}, Int. J. Mod. Phys. conf. Ser. {\bf 14}, 42 (2012).


\bibitem{Rizzo} R. Battesti and C. Rizzo, {\it Magnetic and electric properties of quantum vaccum}, Rept. Prog. in Phys. {\bf 76},  016401 (2013).

\bibitem{Sarazin} X. Sarazin  {\it et al},
{\it Refraction of light by light in vacuum}, Eur. Phys. J. D {\bf 70}, 13 (2016).

\bibitem{Russo2022} J. Russo and P. K. Townsend, {\it Nonlinear Electrodynamics without birefringence}, arXiv : 2211.10689 [hep-th].

\bibitem{Bamber} C. Bamber {\it et al}, {\it Studies of nonlinear QED in collisions of 46.6 GeV electrons with intense laser pulses}, Phys. Rev. D {\bf 60}, 092004 (1999).

\bibitem{Burke} D. L. Burke {\it et al}, {\it Positron Production in Multiphoton Light-by-Light Scattering}, Phys. Rev. Lett. {\bf 79}, 1626 (1997).

\bibitem{Tommasini} D. Tommasini, A. Ferrando, H. Michinel and M. Seco, {\it Precision tests of QED and non-standard models by searching photon-photon scattering in vacuum with high power lasers}, J. High Energy Phys.  {\bf 0911}, 043 (2009).

\bibitem{Ejlli} A.~Ejlli, F.~Della Valle, U.~Gastaldi, G.~Messineo, R.~Pengo, G.~Ruoso and G.~Zavattini,  {\it The PVLAS experiment: A 25 year effort to measure vacuum magnetic birefringence}, Phys. Rept. \textbf{871},  1 (2020).

\bibitem{ATLAS} M. Aaboud  {\it et al} (ATLAS Collaboration), {\it  Evidence for light-by-light scattering in heavy-ion collisions with the ATLAS detector at the LHC}, Nature Physics {\bf 13}, 852 (2017).
%arXiv:1702.01625. Published in {\bf Nature Physics} (2017).

\bibitem{CMS} D. d'Enterria and G. G. da Silveira, {\it Observing Light-by-Light Scattering at the Large Hadron Collider}, Phys. Rev. Lett. {\bf 111}, 080405 (2013); Erratum, Phys. Rev. Lett. {\bf 116}, 129901(E) (2016).

\bibitem{Robertson} S. Robertson  {\it et al, Experiment to observe an optically induced change of the vacuum index}, Phys. Rev. A {\bf 103}, 023524 (2021).

\bibitem{Battesti} R. Battesti {\it et al,  High magnetic fields for fundamental physics}, Phys. Rept. {\bf 765}, 1 (2018).

\bibitem{Ataman} S. Ataman, {\it Vacuum birefringence detection in all-optical scenarios}, Phys. Rev. A  {\bf 97}, 063811 (2018).

\bibitem{Couchot} S. Robertson, A. Mailliet, X. Sarazin, F. Couchot, E. Baynard, J. Demailly,  M. Pittman, A. Djannati-Ata\"\i{}, S. Kazamias, and M. Urban, {\it Experiment to observe an optically induced change of the vacuum index}, Phys. Rev. A {\bf 103}, 023524 (2021).




\bibitem{MJNevesPRD2021} M. J. Neves, Jorge B. de Oliveira, L. P. R. Ospedal and J. A. Helay\"el-Neto, {\it Dispersion Relations in Non-Linear Electrodynamics and the Kinematics of the Compton Effect in a Magnetic Background},  Phys. Rev. D {\bf 104},  015006 (2021).

\bibitem{GEN_BI} P.~Gaete and J.~Helay\"el-Neto,  {\it Remarks on nonlinear Electrodynamics}, Eur.\ Phys.\ J.\ C {\bf 74}, 3182 (2014).


\bibitem{LOG} P.~Gaete and J.~Helay\"el-Neto,  {\it Finite Field-Energy and Interparticle Potential in Logarithmic Electrodynamics}, Eur.\ Phys.\ J.\ C {\bf 74}, 2816 (2014).


%\bibitem{Russo2022} J. Russo and P. K. Townsend, {\it Nonlinear Electrodynamics without birefringence}, arXiv : 2211.10689 [hep-th].


\bibitem{Gaete97} P. Gaete,  {\it On gauge invariant variables in QED}, Z. Phys. C, {\bf 76} 355 (1997).

\bibitem{Sorokin1} I. Bandos, K. Lechner, D. Sorokin and P. T. Townsend,  {\it A non-linear duality-invariant conformal extension of Maxwell's equations}, Phys. Rev. D {\bf 102}, 121703(R) (2020).

\bibitem{Sorokin2} D. Sorokin, {\it Introductory Notes on Non-linear Electrodynamics and its Applications},
 Fortsch. Phys. {\bf 70}, 2200092 (2022).


\bibitem{Sorokin3} K. Lechner, P. Marchetti, A. Sainaghi and D. Sorokin, {\it Maximally symmetric nonlinear extension of electrodynamics and charged particles}, Phys. Rev. D \textbf{106}, 016009 (2022).


\bibitem{Maceda21} D. Flores-Alfonso, B. A. González-Morales, R. Linares, M. Maceda, {\it Black holes and gravitational waves sourced by non-linear duality rotation-invariant conformal electromagnetic matter}, Phys. Lett. B {\bf 812} (2021) 136011.


\bibitem{Bordo21} A. B. Bordo, D. Kubiznak and T. R. Perche, {\it Taub-NUT solutions in conformal electrodynamics}, Phys. Lett. B {\bf 817} (2021) 136312.


\bibitem{Amirabi21} Z. Amirabi and S. H Mazharimousavi, {\it Black-hole solution in nonlinear electrodynamics with the maximum allowable symmetries}, Eur. Phys. J. C {\bf 81} (2021) 207.


\bibitem{Kruglov2022} S.I. Kruglov, {\it Magnetic black holes with generalized ModMax model of nonlinear electrodynamics}, Int. J. Mod. Phys. D {\bf 31} (2022) 04, 2250025.


\bibitem{Barrientos22} J. Barrientos, A. Cisterna, D. Kubiznak and J. Oliva, {\it Accelerated black holes beyond Maxwell's electrodynamics}, Phys. Lett. B {\bf 834} (2022) 137447.


\bibitem{Ali22} Askar Ali and Khalid Saifullah, {\it Charged black holes in 4D Einstein–Gauss–Bonnet gravity coupled to nonlinear electrodynamics with maximum allowable symmetries}, Ann. Phys. {\bf 437} (2022) 168726.


\bibitem{Velni22} H. Babaei-Aghbolagh, K. B. Velni, D. M. Yekta, H. Mohammadzadeh, {\it Marginal $T \bar{T}$-like deformation and modified Maxwell theories in two dimensions}, Phys. Rev. D {\bf 106} (2022) 086022.


\bibitem{Yekta} H. Babaei-Aghbolagh, K. B. Velni, D. M. Yekta, and H. Mohammadzadeh, {\it Emergence of
non-linear electrodynamic theories from $T \bar{T}$-like deformations}, Phys. Lett. B {\bf 829} (2022) 137079.

\bibitem{Conti22} R. Conti, J. Romano and R. Tateo, {\it Metric approach to a $T \bar{T}$-like deformation in arbitrary dimensions}, JHEP {\bf 09} (2022) 085.

\bibitem{Ferko22} C. Ferko, L. Smith and G. Tartaglino-Mazzucchelli, {\it On Current-Squared Flows and ModMax Theories}, SciPost Phys. {\bf 13} (2022) 012.


\bibitem{Nastase2022} Horatiu Nastase, {\it Coupling the precursor of the most general theory of electromagnetism invariant under duality and conformal invariance with scalars, and BIon-type solutions}, Phys. Rev. D {\bf 105} (2022) 105024.


%{\color{red} \bibitem{Sikivie} Pierre Sikivie, {\it Invisible Axion Search Methods}, Rev. Mod. Phys. {\bf 93} (2021) 15004. }

%\bibitem{CondMatter} A. C. Keser, Y. Lyanda-Geller and O. P. Sushkov, {\it Nonlinear Quantum Electrodynamics in Dirac materials}, Phys. Rev. Lett. {\bf 128},  066402 (2022).

\bibitem{Gaete_AHEP_2021} P. Gaete, J.A. Helay\"el-Neto and L.P.R. Ospedal,
{\it Remarks on an anomalous triple gauge boson couplings},
Adv. High Energy Phys. {\bf 2021}, 6621975 (2021).

\bibitem{Gaete_EPJC_2022} M. J. Neves, L. P. R. Ospedal, J.~A.~Helay\"el-Neto and P.~Gaete,
{\it Considerations on anomalous photon and Z-boson self-couplings from the Born\textendash{}Infeld weak hypercharge action},
Eur. Phys. J. C \textbf{82}, 327 (2022).


\bibitem{Plebanski68} J. Plebanski, {\it Lectures on nonlinear Electrodynamics},
Lectures given at the Niels Bohr Institute and NORDITA, October 1968.


\bibitem{Birula70} Z. Bialynicka-Birula and I. Bialynicki-Birula,
{\it Non-linear effects in Quantum Electrodynamics. Photon propagation and photon splitting in an external field}, Phys. Rev. D {\bf 2} (1970) 2341.


\bibitem{Boillat70} G. Boillat, {\it Nonlinear Electrodynamics: Lagrangians and equations of motion}, J. Math. Phys. {\bf 11} (1970) 941.


\bibitem{Boillat72} G. Boillat, {\it Exact plane-wave solution of Born-Infeld Electrodynamics}, Lett. Nuovo Cim. {\bf 4} (1972) 274.


\bibitem{Bialynicki83} I. Bialynicki, {\it Nonlinear Electrodynamics: variations on a theme by Born and Infeld
In Quantum Theory of Particles and Fields}, Birthday Volume dedicated to Jan Lopuszanski
Ed. by B. Jancewicz and J. Lukierski, World Scientific, 1983.


\bibitem{Denisov17} V. I. Denisov, E. E. Dolgaya, V. A. Sokolov and I. P. Denisova,
{\it Conformal invariant vacuum nonlinear Electrodynamics}, Phys. Rev. D {\bf 96} (2017) 036008.


\bibitem{Humblet} J. Humblet, {\it Sur le moment d'impulsion d'une onde éléctromagnetique}, Physica {\bf 10} (1943) 585.


\bibitem{Darwin} C. G. Darwin, {\it Notes n the theory of radiation}, PRSL A {\bf 136} (1932) 36.


\bibitem{Marrucci} L. Marrucci et al., {\it Spin-to-orbital conversion of the angular momentum of light and its classical and quantum implications},
J. Opt. {\bf 13} (2011) 064001.


\bibitem{Stewart} A. M. Stewart, {\it Comparison of different forms for the spin and orbital components of the angular momentum of light},
Int. J. Opt. (2011) 728350.

%\bibitem{MJNevesPRD2021} M. J. Neves, Jorge B. de Oliveira, L. P. R. Ospedal and J. A. Helay\"el-Neto, {\it Dispersion Relations in Non-Linear Electrodynamics and the Kinematics of the Compton Effect in a Magnetic Background},  Phys. Rev. D {\bf 104},  015006 (2021).

\bibitem{Sikivie} Pierre Sikivie, {\it Invisible Axion Search Methods}, Rev. Mod. Phys. {\bf 93} (2021) 15004. 

\bibitem{Gaete_PLB_2004} P. Gaete and C. Wotzasek, {\it On condensation of topological defects and confinement}, Phys. Lett. B \textbf{601}, 108 (2004).

\bibitem{Gaete_PLB_2009} P. Gaete and E. Spallucci, {\it From screening to confinement in a Higgs-like model}, Phys. Lett. B \textbf{675}, 145 (2009).

\bibitem{CondMatter} A. C. Keser, Y. Lyanda-Geller and O. P. Sushkov, {\it Nonlinear Quantum Electrodynamics in Dirac materials}, Phys. Rev. Lett. {\bf 128},  066402 (2022).

%%%%%%%%%%%%%%%%%%%%%%%%%%%%%%%%%%%%%%%
%%%%%%%%%%%%%%%%%%%%%%%%%%%%%%%%%%%%%%%







%
\end{thebibliography}
\end{document}